\documentclass[prx,twocolumn]{revtex4}

\usepackage{graphicx}
\usepackage{dcolumn}
\usepackage{bm}
\usepackage{multirow}
\usepackage{amsmath,amssymb}

\begin{document}


\title{Emergence of electronic modes by doping Kondo insulators in the Kondo lattice and periodic Anderson models}

\author{Masanori Kohno}
\email{KOHNO.Masanori@nims.go.jp}
\affiliation{International Center for Materials Nanoarchitectonics, National Institute for Materials Science, Tsukuba 305-0003, Japan}

\date{\today}

\begin{abstract}
Heavy-fermion or Kondo lattice materials are considered to be typical strongly correlated systems, for which 
mean-field approximations have shown that the Coulomb interaction increases the effective mass and narrows the band gap. 
In this paper, to clarify interaction effects on the electronic excitation, the spectral function of 
the Kondo lattice and periodic Anderson models is studied around Kondo insulators in the strong-coupling regime, 
by using the non-Abelian dynamical density-matrix renormalization group method and perturbation theory. 
Upon doping a Kondo insulator, an electronic mode emerges in the Kondo insulating gap, 
exhibiting the momentum-shifted magnetic dispersion relation. 
Although the ground-state properties are similar to those of a doped band insulator, 
the emergence of the electronic mode reflecting spin-charge separation of the Kondo insulator 
is a crucial interaction effect that allows us to regard the Kondo-insulator--metal transition as a type of Mott transition. 
In addition, electronic modes emerge even in the high-energy regime by doping in the periodic Anderson model. 
These strong-correlation effects have not been expected in conventional mean-field approximations 
and would bring a different perspective on heavy-fermion or Kondo lattice systems. 
\end{abstract}

\maketitle
\section{Introduction} 
The properties of heavy-fermion materials \cite{heavyFermionRMP} have been explained 
by using the Kondo lattice model (KLM) or periodic Anderson model (PAM) \cite{ColemanReview,TsunetsuguRMP,RiceReview}, 
where electrons in conduction orbitals couple with spins or electrons in localized orbitals \cite{Doniach}. 
At half filling, all the electrons in the conduction orbitals couple with those of the localized orbitals, 
and the system becomes an insulator known as a Kondo insulator \cite{ColemanReview,TsunetsuguRMP}. 
The electronic band structure of a Kondo insulator with a spin excitation gap has been considered to 
remain essentially unchanged with or without doping, similar to a conventional band insulator. 
\par
However, unlike a conventional band insulator, the spin gap differs from the charge gap 
in a Kondo insulator \cite{TsunetsuguRMP,spinChargeGapNishino}. 
The spin gap is determined as the energy cost of breaking a spin-singlet state (Kondo singlet) into a spin-triplet state 
in the strong-coupling regime \cite{spingap_1dKLM}, 
whereas the charge gap corresponds to the band gap of electronic excitation. 
Namely, spin-charge separation (difference in the lowest excitation energies between spin and charge) 
occurs \cite{TsunetsuguRMP}. 
Recent theoretical studies on the Mott transition have indicated that, upon doping a Mott insulator 
in which spin-charge separation occurs, an electronic mode emerges 
in the Hubbard gap, exhibiting the magnetic dispersion relation shifted by the Fermi momentum \cite{KohnoRPP,Kohno1DHub,Kohno2DHub,Kohno1DtJ,Kohno2DtJ,KohnoDIS,KohnoAF,KohnoSpin,KohnoHubLadder,KohnoGW}. 
Hence, the key question addressed in this paper is how the electronic excitation behaves near the transition 
to a spin-charge-separated Kondo insulator, 
or how the spin excitation of breaking a Kondo singlet is reflected in the electronic excitation in a doped Kondo insulator. 
\par
To resolve this question, we study the spectral function of the KLM and PAM. 
In this paper, the overall spectral features are clarified by numerical calculations in one dimension, 
and the origins and properties of the characteristic modes are explained in general dimensions 
using effective theory in the strong-coupling regime. 
In the conventional mean-field picture, the increase of the effective mass and the narrowing of the band gap are 
considered to be typical interaction effects in heavy-fermion materials 
\cite{TsunetsuguRMP,RiceReview,RiceUedaPRL,RiceUedaPRB,BrandowPRB,VarmaPRB}. 
In contrast, the numerical and analytical results obtained in this study 
demonstrate that essentially the same characteristic as the Mott transition 
\cite{KohnoRPP,Kohno1DHub,Kohno2DHub,Kohno1DtJ,Kohno2DtJ,KohnoDIS,KohnoAF,KohnoSpin,KohnoHubLadder,KohnoGW} 
appears in a doped Kondo insulator. 
An electronic mode is split into two modes in the low-energy regime by the Coulomb interaction. 
The lower-energy mode gradually loses its spectral weight towards the transition to a Kondo insulator, 
exhibiting the dispersion relation of the spin excitation shifted by the Fermi momentum in the small-doping limit. 
The results imply that the transition to a Kondo insulator can be regarded as a type of Mott transition; 
an electronic mode loses spectral weight because of charge freezing 
but remains dispersing because of the active spin degrees of freedom in the energy regime lower than the charge gap, 
which leads to a spin-charge-separated Kondo insulator. 
\par
Furthermore, in addition to the high-energy modes corresponding to the upper Hubbard band, 
electronic modes emerge even in the high-energy regime upon doping a Kondo insulator in the PAM. 
This implies that doping can affect not only the properties in the vicinity of the Fermi level 
but also those far from the Fermi level 
if hybridization with an orbital away from the Fermi level exists in strongly interacting systems. 
\par
Recently, unconventional features have been observed in Kondo insulator materials \cite{SmB6dHvAScience,SmB6dHvANatPhys,SmB6dHvAiScience,YbB12dHvAJPhysCM,YbB12SdHScience,YbB12heatTransNatPhys,YbB12SdHNatPhys,YbB12HallAnomaly,YbB12SdHQM,SmB6ARPESNatCom,SmB6ARPESPRB,SmB6IngapARPESNatCom,SmB6SpinTextureARPESNatCom}. 
Certain features may be explained in terms of the surface states of 
topological Kondo insulators \cite{DzeroTKIreview,DzeroTKIPRL,TakimotoTKI}. 
This paper shows that even without topological effects, 
electronic features that are unexpected in the conventional mean-field picture 
could arise as a result of strong correlations around Kondo insulators. 
\section{Model and method} 
The KLM and PAM are defined by the following Hamiltonians: 
\begin{align}
\label{eq:KLM}
{\cal H}_{\rm KLM}&=-t\sum_{\langle i,j\rangle,\sigma}\left(c^{\dagger}_{i,\sigma}c_{j,\sigma}+{\mbox {H.c.}}\right)
+J_{\rm K}\sum_{i}{\bm S}^c_{i}\cdot{\bm S}^f_{i}\nonumber\\
&-\mu\sum_{i,\sigma}\left(n^c_{i,\sigma}+n^f_{i,\sigma}\right),\\
\label{eq:PAM}
{\cal H}_{\rm PAM}&=-t\sum_{\langle i,j\rangle,\sigma}\left(c^{\dagger}_{i,\sigma}c_{j,\sigma}+{\mbox {H.c.}}\right)
+U\sum_i n^f_{i,\uparrow}n^f_{i,\downarrow}\nonumber\\
&-t_{\rm K}\sum_{i,\sigma}\left(c^{\dagger}_{i,\sigma}f_{i,\sigma}+{\mbox {H.c.}}\right)-\Delta\sum_{i,\sigma}n^f_{i,\sigma}\nonumber\\
&-\mu\sum_{i,\sigma}\left(n^c_{i,\sigma}+n^f_{i,\sigma}\right), 
\end{align}
where $c_{i,\sigma}$ $(f_{i,\sigma})$ and $n^c_{i,\sigma}$ $(n^f_{i,\sigma})$ denote 
the annihilation and number operators of an electron with spin $\sigma$ 
in the conduction (localized) orbital at site $i$, respectively, 
and ${\bm S}^c_i$ (${\bm S}^f_i$) denotes the spin operator of an electron 
in the conduction (localized) orbital at site $i$. 
Here, $\langle i,j\rangle$ indicates that sites $i$ and $j$ are nearest neighbors. 
Each site has a single conduction orbital and a single localized orbital. 
The electron density and doping concentration are defined as $n=N_{\rm e}/(2L)$ and $\delta=1-n$, respectively, 
where $N_{\rm e}$ and $L$ denote the number of electrons and sites, respectively. 
The doping concentration $\delta$ can also be expressed as $\delta=N_{\rm h}/(2L)$ 
with the number of holes $N_{\rm h}=2L-N_{\rm e}$. 
The magnetization of spin $\sigma$ in units of $\hbar$ is denoted by $s_\sigma$ as follows: 
$s_\sigma=\frac{1}{2}$ for $\sigma=\uparrow$; $s_\sigma=-\frac{1}{2}$ 
for $\sigma=\downarrow$. 
Because the signs of $t$ and $t_{\rm K}$ can be changed by gauge transformations on a bipartite lattice, 
$t>0$ and $t_{\rm K}>0$ are assumed without loss of generality unless otherwise mentioned. 
In the KLM, each localized orbital is assumed to be singly occupied by an electron. 
The KLM can be derived as an effective model of the PAM in the large-$U$ and large-$\Delta$ regime \cite{SWtrans,TsunetsuguRMP}. 
\par
We study the spectral function $A({\bm k},\omega)$ and dynamical spin structure factor $S({\bm k},\omega)$ 
at zero temperature on a $d$-dimensional cubic lattice defined as 
\begin{equation}
\label{eq:addAkw}
A({\bm k},\omega)=\frac{1}{2}\sum_{l,\sigma}\left(|\langle l|c^{\dagger}_{{\bm k},\sigma}|{\rm GS}\rangle|^2
+|\langle l|f^{\dagger}_{{\bm k},\sigma}|{\rm GS}\rangle|^2\right)\delta(\omega-\varepsilon_l)
\end{equation}
for $\omega>0$, 
\begin{equation}
\label{eq:remAkw}
A({\bm k},\omega)=\frac{1}{2}\sum_{l,\sigma}\left(|\langle l|c_{{\bm k},\sigma}|{\rm GS}\rangle|^2
+|\langle l|f_{{\bm k},\sigma}|{\rm GS}\rangle|^2\right)\delta(\omega+\varepsilon_l)
\end{equation}
for $\omega<0$, and 
\begin{equation}
\label{eq:Skw}
S({\bm k},\omega)=\frac{1}{6}\sum_{l,\alpha}
|\langle l|(S^{c,\alpha}_{\bm k}-S^{f,\alpha}_{\bm k})|{\rm GS}\rangle|^2\delta(\omega-\varepsilon_l), 
\end{equation}
where $c^{\dagger}_{{\bm k},\sigma}$, $f^{\dagger}_{{\bm k},\sigma}$, $S^{c,\alpha}_{\bm k}$, 
and $S^{f,\alpha}_{\bm k}$ denote the Fourier transform of 
$c^{\dagger}_{i,\sigma}$, $f^{\dagger}_{i,\sigma}$, $S^{c,\alpha}_{i}$, and $S^{f,\alpha}_{i}$, respectively 
($\alpha=x$, $y$, and $z$). 
Here, $|{\rm GS}\rangle$ and $|l\rangle$ represent the ground state 
and an eigenstate with excitation energy $\varepsilon_l$ from $|{\rm GS}\rangle$, respectively. 
In the KLM, the contributions from the localized orbitals in Eqs. (\ref{eq:addAkw}) and (\ref{eq:remAkw}) are zero, 
because an electron cannot be added to or removed from the localized orbitals. 
In the KLM and PAM at $U=2\Delta$, $A({\bm k},\omega)$ in an electron-doped system can be obtained 
as $A({\bm k}+{\bm \pi},-\omega)$ in a hole-doped system using particle-hole and gauge transformations. 
\par
In this paper, the numerical results for $A({\bm k},\omega)$ and $S({\bm k},\omega)$ obtained 
using the non-Abelian dynamical density-matrix renormalization group (DDMRG) 
method \cite{KohnoDIS,Kohno1DtJ,Kohno2DtJ,KohnoHubLadder,nonAbelianHub,nonAbeliantJ,nonAbelianThesis,DDMRG} 
in the one-dimensional (1D) KLM with $L=40$ and 
those in the 1D PAM with $L=30$ under open boundary conditions are shown. 
In the non-Abelian DDMRG calculations, 240 eigenstates of the density matrix were retained. 
The ground state was assumed to be a spin singlet in a system with an even number of electrons. 
\section{Spectral features} 
\begin{figure*}
\includegraphics[width=\linewidth]{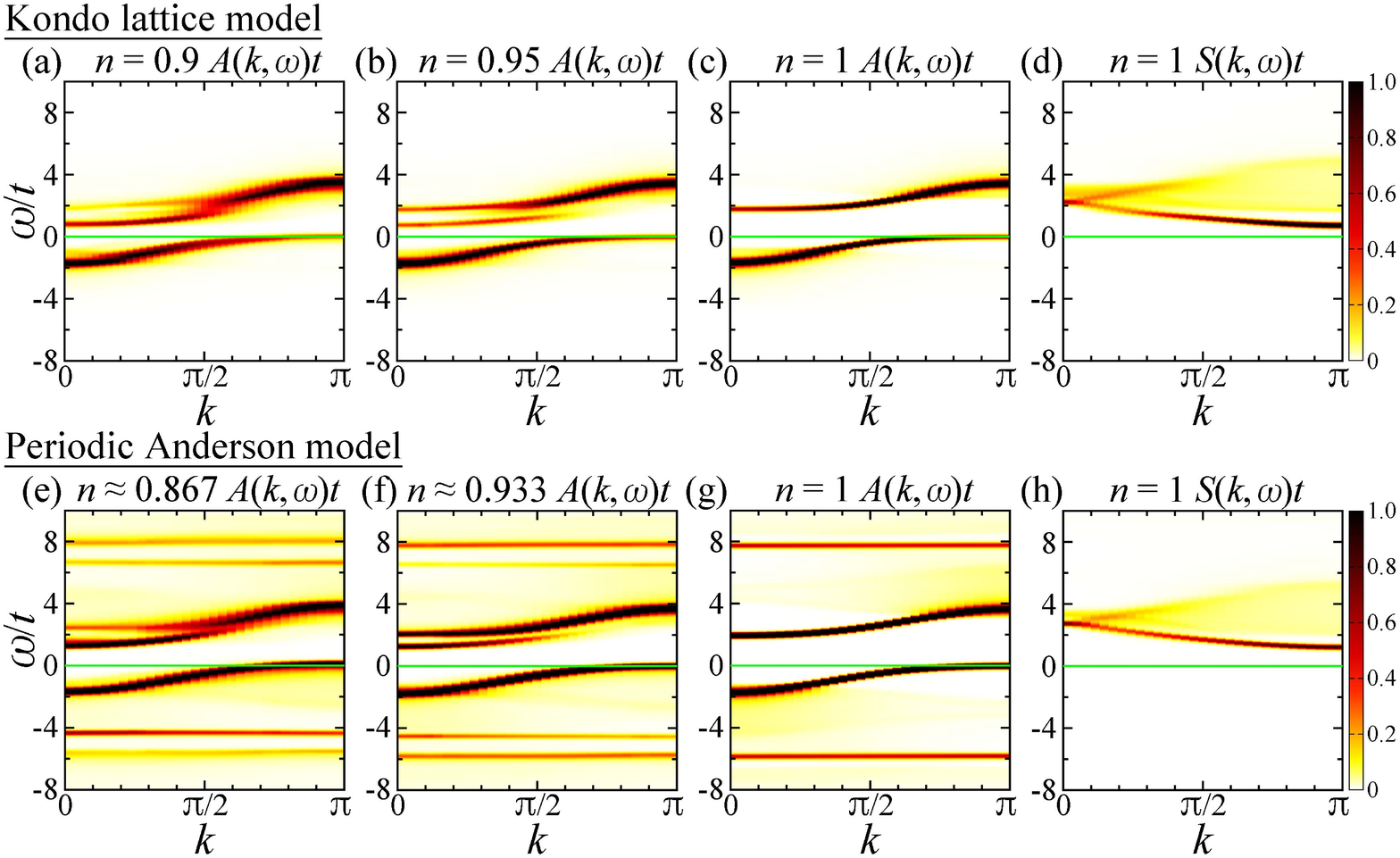}
\caption{Spectral function $A(k,\omega)$ and dynamical spin structure factor $S(k,\omega)$ 
of the 1D KLM for $J_{\rm K}=2t$ [(a)--(d)] and PAM for $U=8t$, $t_{\rm K}=2t$, and $\Delta=4t$ [(e)--(h)] 
obtained using the non-Abelian DDMRG method. 
(a)--(c) $A(k,\omega)t$ of the 1D KLM at $n=0.9$ [(a)], $n=0.95$ [(b)], and $n=1$ [(c)]. 
(d) $S(k,\omega)t$ of the 1D KLM at $n=1$. 
(e)--(g) $A(k,\omega)t$ of the 1D PAM at $n\approx 0.867$ [(e)], $n\approx 0.933$ [(f)], and $n=1$ [(g)]. 
(h) $S(k,\omega)t$ of the 1D PAM at $n=1$. The green lines indicate $\omega=0$ [Fermi level in (a)--(c) and (e)--(g)]. 
Gaussian broadening with a standard deviation of $0.1t$ was used.}
\label{fig:Akw}
\end{figure*}
The numerical results for the 1D KLM and PAM are shown in Fig. \ref{fig:Akw}. 
In a Kondo insulator at half filling ($n=1$) \cite{TsunetsuguRMP,ColemanReview,spinChargeGapNishino}, 
there are two dominant modes around the Fermi level [Figs. \ref{fig:Akw}(c) and \ref{fig:Akw}(g)]. 
They are separated by the Kondo insulating gap. 
In the PAM, almost flat modes exist in the high-$|\omega|$ regime 
for both $\omega>0$ and $\omega<0$ [Fig. \ref{fig:Akw}(g)]. 
\par
On doping the Kondo insulator, a mode emerges in the gap [Figs. \ref{fig:Akw}(b) and \ref{fig:Akw}(f)]. 
As the doping concentration increases, the emergent mode gradually gains spectral weight 
and merges with the dominant mode for $\omega>0$ [Figs. \ref{fig:Akw}(a) and \ref{fig:Akw}(e)]. 
In the PAM, in addition to the in-gap mode, additional high-$|\omega|$ modes emerge 
for both $\omega>0$ and $\omega<0$ [Fig. \ref{fig:Akw}(f)]. 
The emergent modes gradually gain spectral weight as the doping concentration increases [Fig. \ref{fig:Akw}(e)]. 
\par
Usually, a doped Kondo insulator is considered to exhibit essentially the same band structure 
as that of a Kondo insulator, similar to a doped band insulator. 
However, the numerical results [Figs. \ref{fig:Akw}(a)--\ref{fig:Akw}(c), \ref{fig:Akw}(e)--\ref{fig:Akw}(g)] indicate that 
an in-gap mode emerges on doping a Kondo insulator. 
In addition, high-$|\omega|$ modes emerge on doping a Kondo insulator in the PAM 
even though they are far from the Fermi level [Figs. \ref{fig:Akw}(e)--\ref{fig:Akw}(g)]. 
These doping-induced modes gain spectral weight as the doping concentration increases, 
and become dominant in the large-doping regime. 
This remarkable characteristic has not been expected in the conventional mean-field picture 
\cite{ColemanReview,TsunetsuguRMP,RiceReview,RiceUedaPRL,RiceUedaPRB,BrandowPRB,VarmaPRB}. 
The origins and properties of these modes are explained from a strong-coupling viewpoint in this paper. 
\section{Kondo insulator} 
\subsection{$t=0$} 
\begin{table*}
\caption{Eigenstates and energies on a single site.}
\begin{tabular}{wc{1cm}wc{1cm}wc{1cm}wc{3.6cm}wc{3.6cm}wc{3.6cm}}
\hline\hline
$N_{\rm e}$&$S$&$S_z$&KLM&PAM for $U=\infty$&PAM for $U=2\Delta$\\\hline 
\multirow{2}{*}{0}&\multirow{2}{*}{0}&\multirow{2}{*}{0}&\multirow{2}{*}{---}&$|0\rangle$&$|0\rangle$\\ 
&&&&0&0\\\hline 
\multirow{2}{*}{1}&\multirow{2}{*}{$\frac{1}{2}$}&\multirow{2}{*}{$s_\sigma$}&$|1_{\sigma}\rangle$
&$|1_{\sigma}^{\pm}\rangle$&$|1_{\sigma}^{\pm}\rangle$\\ 
&&&$-\mu$&$E^{\pm}_{t_{\rm K}}-\mu$&$E^{\pm}_{t_{\rm K}}-\mu$\\\hline 
\multirow{8}{*}{2}&\multirow{4}{*}{1}&\multirow{2}{*}{$2s_\sigma$}
&$|{\rm T}_{2s_\sigma}\rangle$&$|{\rm T}_{2s_\sigma}\rangle$&$|{\rm T}_{2s_\sigma}\rangle$\\ 
&&&$\frac{J_{\rm K}}{4}-2\mu$&$-\Delta-2\mu$&$-\Delta-2\mu$\\\cline{3-6}
&&\multirow{2}{*}{0}&$|{\rm T}_0\rangle$&$|{\rm T}_0\rangle$&$|{\rm T}_0\rangle$\\
&&&$\frac{J_{\rm K}}{4}-2\mu$&$-\Delta-2\mu$&$-\Delta-2\mu$\\\cline{2-6}
&\multirow{4}{*}{0}&\multirow{4}{*}{0}
&$|{\rm S}\rangle$
&$|\psi^{\pm}_{\infty}\rangle$
&$|\psi^{\pm}_{2\Delta}\rangle$\\ 
&&&$-\frac{3J_{\rm K}}{4}-2\mu$
&$E^{\pm}_{\sqrt{2}t_{\rm K}}-2\mu$&$E^{\pm}_{2t_{\rm K}}-2\mu$\\\cline{4-6} 
&&&\multirow{2}{*}{---}&\multirow{2}{*}{---}&$|{\rm D}^-\rangle$\\ 
&&&&&$-2\mu$\\\hline 
\multirow{2}{*}{3}&\multirow{2}{*}{$\frac{1}{2}$}&\multirow{2}{*}{$s_\sigma$}&$|3_{\sigma}\rangle$&$|3_{\sigma}\rangle$
&$|3_{\sigma}^{\pm}\rangle$\\ 
&&&$-3\mu$&$-\Delta-3\mu$&$E^{\pm}_{t_{\rm K}}-3\mu$\\\hline 
\multirow{2}{*}{4}&\multirow{2}{*}{0}&\multirow{2}{*}{0}&\multirow{2}{*}{---}&\multirow{2}{*}{---}&$|4\rangle$\\ 
&&&&&$-4\mu$\\\hline\hline 
\end{tabular}
\footnote[0]{
\hspace{1.3cm}\begin{tabular}{p{3cm}p{3cm}p{3cm}p{3cm}}
$|0\rangle=|0,0\rangle$,&$|4\rangle=|\uparrow\downarrow,\uparrow\downarrow\rangle$,
&$|1_{\sigma}\rangle=|0,\sigma\rangle$,&$|3_{\sigma}\rangle=|\uparrow\downarrow,\sigma\rangle$,\\
\multicolumn{2}{l}{$|1_{\sigma}^{\pm}\rangle=u^{\mp}_{t_{\rm K}}|0,\sigma\rangle\mp u^{\pm}_{t_{\rm K}}|\sigma,0\rangle$,}&
\multicolumn{2}{l}{$|3_{\sigma}^{\pm}\rangle=u^{\mp}_{t_{\rm K}}|\uparrow\downarrow,\sigma\rangle\pm u^{\pm}_{t_{\rm K}}|\sigma,\uparrow\downarrow\rangle$,}\\
\multicolumn{2}{l}{$|{\rm T}_{2s_\sigma}\rangle=|\sigma,\sigma\rangle$,}&
\multicolumn{2}{l}{$|{\rm T}_0\rangle=\frac{1}{\sqrt{2}}\left(|\uparrow,\downarrow\rangle+|\downarrow,\uparrow\rangle\right)$,}\\
\multicolumn{2}{l}{$|{\rm S}\rangle=\frac{1}{\sqrt{2}}\left(|\uparrow,\downarrow\rangle-|\downarrow,\uparrow\rangle\right)$,}&
\multicolumn{2}{l}{$|{\rm D}^{\pm}\rangle=\frac{1}{\sqrt{2}}(|0,\uparrow\downarrow\rangle\pm|\uparrow\downarrow,0\rangle)$,}\\
\multicolumn{2}{l}{$|\psi^{\pm}_{\infty}\rangle=u^{\mp}_{\sqrt{2}t_{\rm K}}|{\rm S}\rangle\mp u^{\pm}_{\sqrt{2}t_{\rm K}}|\uparrow\downarrow,0\rangle$,}&
\multicolumn{2}{l}{$|\psi^{\pm}_{2\Delta}\rangle=u^{\mp}_{2t_{\rm K}}|{\rm S}\rangle\mp u^{\pm}_{2t_{\rm K}}|{\rm D}^+\rangle$.}\\
\multicolumn{2}{l}{$E^{\pm}_x=-\frac{\Delta}{2}\pm\frac{\sqrt{\Delta^2+4x^2}}{2}$.}&
\multicolumn{2}{l}{$u^{\pm}_x=\sqrt{\frac{1}{2}\left(1\pm\frac{\Delta}{\sqrt{\Delta^2+4x^2}}\right)}$.}\\
\multicolumn{4}{l}{The single-site state with conduction-orbital state $\alpha_1$ and 
localized-orbital state $\alpha_2$ is represented by $|\alpha_1,\alpha_2\rangle$.}
\end{tabular}}
\label{tbl:1site}
\end{table*}
To clarify the origins of the doping-induced modes, we consider the small-$t$ regime 
($t\ll J_{\rm K}$, $t_{\rm K}$, $\Delta$, and $U$). 
Without intersite hopping ($t=0$), the system is decomposed into isolated sites. 
The eigenstates on a site are classified by the number of electrons $N_{\rm e}$, spin $S$, and magnetization $S_z$ 
as listed in Table \ref{tbl:1site} \cite{OneSiteKondo}. 
Hereafter, the energy of single-site state $\alpha$ is denoted as ${\bar E}_{\alpha}$. 
\subsection{Modes in the KLM} 
\label{sec:modesKLM}
When intersite hopping ($t$) is turned on, the single-site states are coupled with those of the neighboring sites. 
In the small-$t$ regime, effective eigenstates are obtained using perturbation theory with respect to $t$. 
\par
In this subsection, we shortly review the properties of the spin and electronic modes 
of a Kondo insulator in the small-$t/J_{\rm K}$ limit of the KLM obtained in Ref. \cite{TsunetsuguRMP}. 
By using the single-site ground state $|{\rm S}\rangle_j$ at site $j$ (Table \ref{tbl:1site}), 
the ground state can be effectively expressed as 
\begin{equation}
|{\rm GS}\rangle^{\rm KLM}=\prod_{j=1}^{L}|{\rm S}\rangle_j, 
\label{eq:GSKLM}
\end{equation}
whose energy is obtained as 
\begin{equation}
E_{{\rm GS},N_{\rm h}=0}^{\rm KLM}=\left(-\frac{3J_{\rm K}}{4}-2\mu\right)L+d\xi_{\rm SS}L
\end{equation}
up to $\mathcal{O}(t^2/J_{\rm K})$. 
Here, $\xi_{XX^{\prime}}$ denotes the bond energy between $|X\rangle_i$ and $|X^{\prime}\rangle_j$ 
on neighboring sites $i$ and $j$ obtained in the second-order perturbation theory. 
\par
The spin excited states can be expressed as 
\begin{equation}
|\alpha\rangle_{\bm k}^{\rm KLM}=\frac{1}{\sqrt{L}}\sum_{j=1}^L\mathrm{e}^{i{\bm k}\cdot{\bm r}_j}
|\alpha\rangle_j\prod_{l\ne j}^{L}|{\rm S}\rangle_l
\label{eq:modeKLM}
\end{equation}
for $\alpha={\rm T}_1$, ${\rm T}_{-1}$, or ${\rm T}_0$ (Table \ref{tbl:1site}), 
whose excitation energies from $|{\rm GS}\rangle^{\rm KLM}$ are obtained as 
\begin{equation}
e_{\rm spin}^{\rm KLM}({\bm k})=J_{\rm eff}^{\rm KLM}d\gamma({\bm k})+J_{\rm K}+2d(\xi_{\alpha S}-\xi_{\rm SS}) 
\label{eq:spinEneKLM}
\end{equation}
up to $\mathcal{O}(t^2/J_{\rm K})$ with 
\begin{equation}
\gamma({\bm k})=\frac{1}{d}\sum_{i=1}^d \cos k_i, 
\end{equation}
where $k_x$, $k_y$, and $k_z$ are denoted as $k_i$ for $i$=1, 2, and 3, respectively. 
In the second-order perturbation theory, 
\begin{equation}
J_{\rm eff}^{\rm KLM}=\frac{4t^2}{J_{\rm K}},\quad\xi_{\rm SS}=
-\frac{2t^2}{3J_{\rm K}},\quad\xi_{\alpha{\rm S}}=-\frac{2t^2}{J_{\rm K}} 
\end{equation}
for $\alpha={\rm T}_1$, ${\rm T}_{-1}$, and ${\rm T}_0$ \cite{TsunetsuguRMP}. 
\par
\begin{figure}
\includegraphics[width=\linewidth]{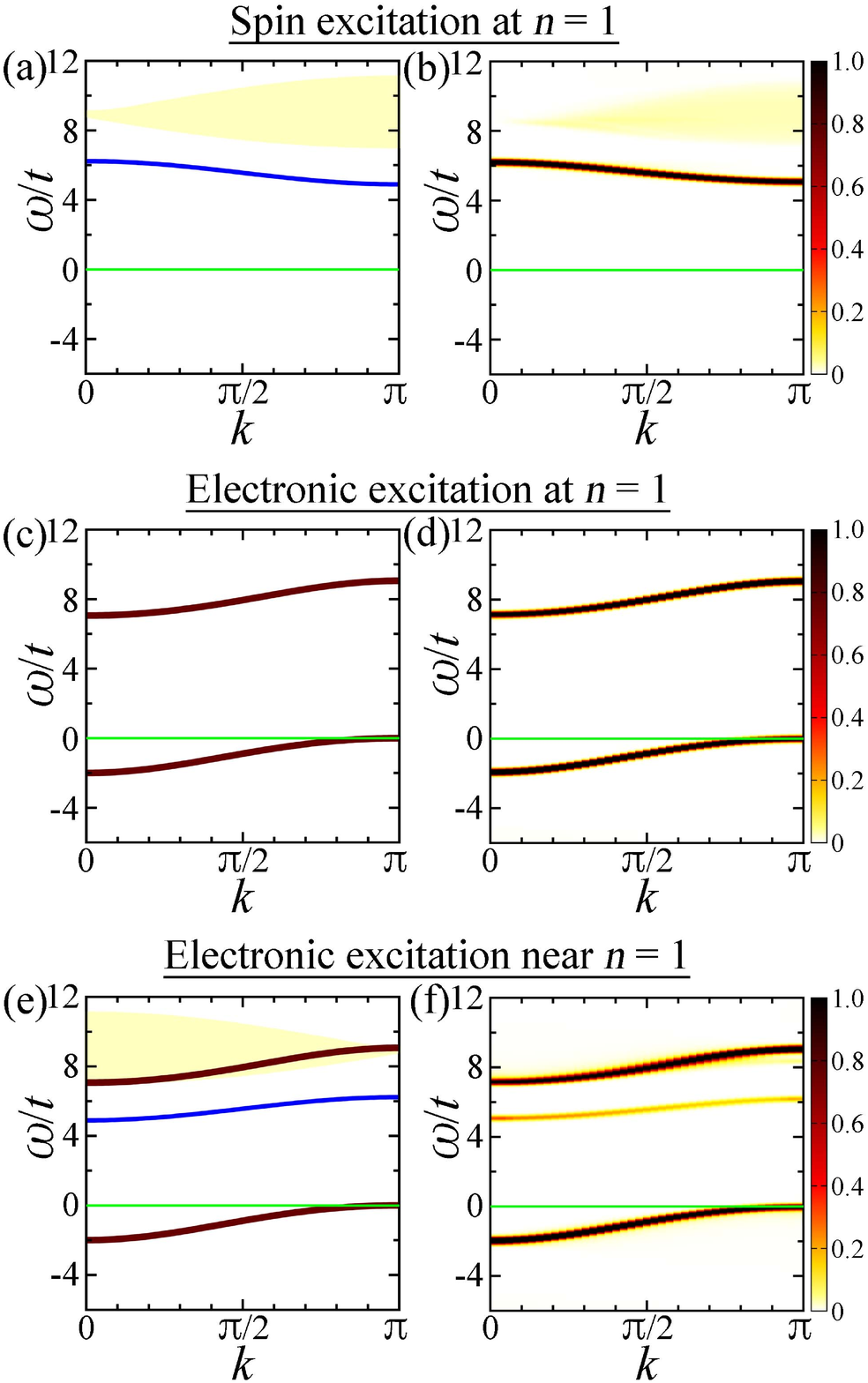}
\caption{Excitation spectra, dynamical spin structure factor $S(k,\omega)$, and spectral function $A(k,\omega)$ 
of the 1D KLM for $J_{\rm K}=6t$. 
(a) Spin excitation spectrum at $n=1$: 
$\omega=e_{\rm ph}^{\rm KLM}(k;p)$ (light-yellow region), 
$\omega=e_{\rm spin}^{\rm KLM}(k)$ (solid blue curve). 
(b) $S(k,\omega)t$ at $n=1$. 
(c) Electronic excitation spectrum at $n=1$: 
$\omega=\epsilon_{3_{\sigma}}^{\rm KLM}(k)$ (solid brown curve for $\omega>0$), 
$\omega=-\epsilon_{1_{\sigma}}^{\rm KLM}(k)$ (solid brown curve for $\omega<0$). 
(d) $A(k,\omega)t$ at $n=1$. 
(e) Electronic excitation spectrum for $n\rightarrow 1$: 
$\omega=e_{\rm ph}^{\rm KLM}(k-\pi;p)$ (light-yellow region), 
$\omega=\epsilon_{3_{\sigma}}^{\rm KLM}(k)$ (solid brown curve for $\omega>0$), 
$\omega=e_{\rm spin}^{\rm KLM}(k-\pi)$ (solid blue curve), 
$\omega=-\epsilon_{1_{\sigma}}^{\rm KLM}(k)$ (solid brown curve for $\omega<0$). 
(f) $A(k,\omega)t$ at $n=0.975$. 
The green lines indicate $\omega=0$ [Fermi level in (c)--(f)]. 
Gaussian broadening with a standard deviation of $0.1t$ was used for the numerical results obtained 
using the non-Abelian DDMRG method [(b), (d), and (f)].}
\label{fig:weff}
\end{figure}

The dispersion relation of the spin excitation mode $\omega=e_{\rm spin}^{\rm KLM}({\bm k})$ is shown 
by the blue curve in Fig. \ref{fig:weff}(a) for $J_{\rm K}/t=6$ in the 1D KLM. 
The validity of the perturbation theory is confirmed using the non-Abelian DDMRG method [Fig. \ref{fig:weff}(b)]. 
\par
The electronic excited states can also be expressed as Eq. (\ref{eq:modeKLM}) 
with $\alpha=1_{\sigma}$ and $3_{\sigma}$ (Table \ref{tbl:1site}), 
whose excitation energies are obtained as \cite{TsunetsuguRMP} 
\begin{equation}
\epsilon_{\alpha}^{\rm KLM}({\bm k})=-2t_{\rm eff}^{\rm KLM}d\gamma({\bm k})+\frac{3J_{\rm K}}{4}
+2d(\xi_{\alpha{\rm S}}-\xi_{\rm SS})+\eta_{3{\rm site}}^{\rm KLM}({\bm k})-\mu^{\prime}
\label{eq:eleEneKLM}
\end{equation}
up to $\mathcal{O}(t^2/J_{\rm K})$, where $t_{\rm eff}^{\rm KLM}=-t/2$ and $\mu^{\prime}=-\mu$ for $\alpha=1_{\sigma}$; 
$t_{\rm eff}^{\rm KLM}=t/2$ and $\mu^{\prime}=\mu$ for $\alpha=3_{\sigma}$, 
\begin{equation}
\xi_{\alpha{\rm S}}=-\frac{3t^2}{4J_{\rm K}}
\end{equation}
for $\alpha=1_{\sigma}$ and $3_{\sigma}$, and the three-site hopping energy \cite{TsunetsuguRMP} 
\begin{equation}
\eta_{3{\rm site}}^{\rm KLM}({\bm k})=\frac{t^2d}{3J_{\rm K}}[2d\gamma({\bm k})^2-1]. 
\end{equation}
\par
The dispersion relations of the electronic modes 
$\omega=\epsilon_{3_\sigma}^{\rm KLM}({\bm k})$ and 
$\omega=-\epsilon_{1_\sigma}^{\rm KLM}({\bm k})$ are shown 
by the brown curves in Fig. \ref{fig:weff}(c) for $J_{\rm K}/t=6$ in the 1D KLM, 
where the Fermi level is tuned to the top of the lower band. 
The validity of the perturbation theory is confirmed using the non-Abelian DDMRG method [Fig. \ref{fig:weff}(d)]. 
\subsection{Modes in the PAM} 
\label{sec:modesPAM}
The above perturbation theory can also be applied to the PAM in the small-$t$ limit. 
In the PAM, the single-site ground states $|\psi_U^-\rangle_j$ for $U=2\Delta$ and $\infty$ (Table \ref{tbl:1site}) are 
used instead of $|{\rm S}\rangle_j$ in Eq. (\ref{eq:GSKLM}): 
\begin{equation}
|{\rm GS}\rangle^{\rm PAM}=\prod_{j=1}^{L}|\psi_U^-\rangle_j,
\end{equation}
whose energy is obtained as 
\begin{equation}
E_{{\rm GS},N_{\rm h}=0}^{\rm PAM}={\bar E}_{\psi_U^-}L+d\xi_{\psi_U^-\psi_U^-}L
\end{equation}
up to $\mathcal{O}(t^2/\Delta{\bar E})$, where $\Delta{\bar E}$ denotes excitation energy to an intermediate state 
in the second-order perturbation theory. 
Here, ${\bar E}_{\psi_{2\Delta}^-}=E_{2t_{\rm K}}^--2\mu$; 
${\bar E}_{\psi_{\infty}^-}=E_{\sqrt{2}t_{\rm K}}^--2\mu$ (Table \ref{tbl:1site}) with 
\begin{equation}
E^{\pm}_x=-\frac{\Delta}{2}\pm\frac{\sqrt{\Delta^2+4x^2}}{2}.
\end{equation} 
\par
Similarly to Eq. (\ref{eq:modeKLM}), the excited states for spin and electronic modes can be expressed as 
\begin{equation}
|\alpha\rangle_{\bm k}^{\rm PAM}=\frac{1}{\sqrt{L}}\sum_{j=1}^L\mathrm{e}^{i{\bm k}\cdot{\bm r}_j}
|\alpha\rangle_j\prod_{l\ne j}^{L}|\psi_U^-\rangle_l, 
\label{eq:modePAM}
\end{equation}
by using single-site state $|\alpha\rangle$ (Table \ref{tbl:1site}). 
\par
The spin excited states are obtained as Eq. (\ref{eq:modePAM}) with $\alpha={\rm T}_1$, ${\rm T}_{-1}$, or ${\rm T}_0$. 
The excitation energies from $|{\rm GS}\rangle^{\rm PAM}$ are obtained similarly to Eq. (\ref{eq:spinEneKLM}) as 
\begin{equation}
\label{eq:spinEnePAM}
e_{\rm spin}^{\rm PAM}({\bm k})=J_{\rm eff}^{\rm PAM}d\gamma({\bm k})+{\bar E}_{\alpha}-{\bar E}_{\psi_U^-}
+2d(\xi_{\alpha\psi_U^-}-\xi_{\psi_U^-\psi_U^-}) 
\end{equation}
up to $\mathcal{O}(t^2/\Delta{\bar E})$, where ${\bar E}_{\alpha}=-\Delta-2\mu$ (Table \ref{tbl:1site}). 
Here, $J_{\rm eff}^{\rm PAM}$, $\xi_{\alpha\psi_U^-}$, and $\xi_{\psi_U^-\psi_U^-}$ are 
quantities of $\mathcal{O}(t^2/\Delta{\bar E})$ obtained in the second-order perturbation theory. 
\par
The electronic excited states are obtained as Eq. (\ref{eq:modePAM}) with 
$\alpha=1_{\sigma}^\pm$ and $3_{\sigma}^{\pm}$ for $U=2\Delta$; 
$\alpha=1_{\sigma}^\pm$ and $3_{\sigma}$ for $U=\infty$. 
The excitation energies are obtained similarly to Eq. (\ref{eq:eleEneKLM}) as 
\begin{align}
\epsilon_{\alpha}^{\rm PAM}({\bm k})&=-2t_{{\rm eff},\alpha}^{\rm PAM}d\gamma({\bm k})
+{\bar E}_{\alpha}-{\bar E}_{\psi_U^-}\nonumber\\
&+2d(\xi_{\alpha\psi_U^-}-\xi_{\psi_U^-\psi_U^-})+\eta_{3{\rm site},\alpha}^{\rm PAM}({\bm k})
\label{eq:eleEnePAM}
\end{align}
up to $\mathcal{O}(t^2/\Delta{\bar E})$, 
where $t_{{\rm eff},\alpha}^{\rm PAM}$ is the effective hopping integral of $\mathcal{O}(t)$, 
whose sign is the same as that of the KLM case, obtained in the first-order perturbation theory, 
and $\xi_{\alpha\psi_U^-}$, $\xi_{\psi_U^-\psi_U^-}$, and $\eta_{3{\rm site},\alpha}^{\rm PAM}({\bm k})$ are quantities of 
$\mathcal{O}(t^2/\Delta{\bar E})$ obtained in the second-order perturbation theory. 
\subsection{Identification of modes} 
From the viewpoint of small $t$, the spectral features at $n=1$ 
in Figs. \ref{fig:Akw}(c), \ref{fig:Akw}(d), \ref{fig:Akw}(g), and \ref{fig:Akw}(h) can be explained as follows. 
The spin excitation mode of the KLM in Fig. \ref{fig:Akw}(d) and that of the PAM in Fig. \ref{fig:Akw}(h) 
can be effectively identified with those of $\alpha={\rm T}_1$, ${\rm T}_{-1}$, or ${\rm T}_0$ in Eq. (\ref{eq:modeKLM}) 
and Eq. (\ref{eq:modePAM}), respectively. 
The electronic modes of the KLM in Fig. \ref{fig:Akw}(c) can be effectively identified with 
the mode of $\alpha=1_{\sigma}$ for $\omega<0$ and that of $\alpha=3_{\sigma}$ for $\omega>0$ in Eq. (\ref{eq:modeKLM}). 
The dominant electronic modes of the PAM around the Fermi level in Fig. \ref{fig:Akw}(g) can be effectively identified with 
the mode of $\alpha=1_{\sigma}^-$ for $\omega<0$ and that of $\alpha=3_{\sigma}^-$ for $\omega>0$ 
in Eq. (\ref{eq:modePAM}). 
\par
In the PAM, in addition to the dominant electronic modes around the Fermi level, 
almost flat modes exist in the high-$|\omega|$ regime for both $\omega<0$ and $\omega>0$ in Fig. \ref{fig:Akw}(g). 
These modes can be effectively identified with the mode of $\alpha=1_{\sigma}^+$ for $\omega<0$ 
and that of $\alpha=3_{\sigma}^+$ for $\omega>0$ in Eq. (\ref{eq:modePAM}). 
Thus the reason for the existence of not only the low-$|\omega|$ dominant modes 
but also the high-$|\omega|$ electronic modes in the PAM can be understood 
in terms of the existence of two $N_{\rm e}=1$ states ($|1_{\sigma}^\pm\rangle$) 
and two $N_{\rm e}=3$ states ($|3_{\sigma}^\pm\rangle$) per spin in the single-site PAM 
for $U=2\Delta$ in contrast to the single-site KLM which has one $N_{\rm e}=1$ state ($|1_{\sigma}\rangle$) 
and one $N_{\rm e}=3$ state ($|3_{\sigma}\rangle$) per spin (Table \ref{tbl:1site}). 
\par
Furthermore, the origin of the high-$|\omega|$ mode characterized by $|1_{\sigma}^+\rangle$ ($|3_{\sigma}^+\rangle$) 
can be understood primarily as the removed (added) electron in the localized orbital. 
This is because the $N_{\rm e}=1$ and $N_{\rm e}=3$ single-site states are expressed as 
\begin{align}
|1_{\sigma}^{\pm}\rangle&=u^{\mp}_{t_{\rm K}}|0,\sigma\rangle\mp u^{\pm}_{t_{\rm K}}|\sigma,0\rangle,\\
|3_{\sigma}^{\pm}\rangle&=u^{\mp}_{t_{\rm K}}|\uparrow\downarrow,\sigma\rangle
\pm u^{\pm}_{t_{\rm K}}|\sigma,\uparrow\downarrow\rangle,
\end{align}
where $|\alpha_1,\alpha_2\rangle$ represents the single-site state 
with conduction-orbital state $\alpha_1$ and localized-orbital state $\alpha_2$, and 
\begin{equation}
u^{\pm}_{t_{\rm K}}=\sqrt{\frac{1}{2}\left(1\pm\frac{\Delta}{\sqrt{\Delta^2+4t_{\rm K}^2}}\right)}
\label{eq:ux}
\end{equation}
 (Table \ref{tbl:1site}). Here, $u^+_{t_{\rm K}} > u^-_{t_{\rm K}}$ for $\Delta>0$; 
$u^+_{t_{\rm K}}=\mathcal{O}(1)$ and $u^-_{t_{\rm K}}=\mathcal{O}(t_{\rm K}/\Delta)$ for $t_{\rm K}\ll\Delta$. 
On the other hand, the low-$|\omega|$ mode characterized by $|1_{\sigma}^-\rangle$ ($|3_{\sigma}^-\rangle$) 
is expected to have a considerable conduction-orbital component.
\subsection{Spin excitation continuum} 
To explain the continuum above the dominant mode in the spin excitation spectrum 
[Figs. \ref{fig:Akw}(d) and \ref{fig:Akw}(h)], we consider the following particle-hole excited states: 
\begin{align}
\label{eq:contKLM}
|\alpha_1\alpha_2\rangle_{{\bm k};{\bm p}}^{\rm KLM}=&\frac{1}{\sqrt{L(L-1)}}\sum_{i\ne j}^L
\mathrm{e}^{i({\bm k}-{\bm p})\cdot{\bm r}_i}\mathrm{e}^{i{\bm p}\cdot{\bm r}_j}\nonumber\\
&|\alpha_1\rangle_i|\alpha_2\rangle_j\prod_{l\ne i,j}^{L}|{\rm S}\rangle_l,\\
|\alpha_1\alpha_2\rangle_{{\bm k};{\bm p}}^{\rm PAM}=&\frac{1}{\sqrt{L(L-1)}}\sum_{i\ne j}^L
\mathrm{e}^{i({\bm k}-{\bm p})\cdot{\bm r}_i}\mathrm{e}^{i{\bm p}\cdot{\bm r}_j}\nonumber\\
&|\alpha_1\rangle_i|\alpha_2\rangle_j\prod_{l\ne i,j}^{L}|\psi_U^-\rangle_l
\label{eq:contPAM}
\end{align}
with $\alpha_1=3_{\sigma}$ and $\alpha_2=1_{\sigma^\prime}$ for the KLM; 
$\alpha_1=3_{\sigma}^-$ and $\alpha_2=1_{\sigma^\prime}^-$ for the $U=2\Delta$ PAM; 
$\alpha_1=3_{\sigma}$ and $\alpha_2=1_{\sigma^\prime}^-$ for the $U=\infty$ PAM. 
Here, $(\sigma,\sigma^\prime)=(\uparrow,\uparrow)$ for $S_z=1$; 
$(\sigma,\sigma^\prime)=(\downarrow,\downarrow)$ for $S_z=-1$; 
$(\sigma,\sigma^\prime)=(\uparrow,\downarrow)$ and $(\downarrow,\uparrow)$ for $S_z=0$. 
The excitation energies can be approximately obtained as 
\begin{align}
\label{eq:EphKLM}
e_{\rm ph}^{\rm KLM}({\bm k};{\bm p})
=\epsilon_{\alpha_1}^{\rm KLM}({\bm k}-{\bm p})+\epsilon_{\alpha_2}^{\rm KLM}({\bm p}),\\ 
\label{eq:EphPAM}
e_{\rm ph}^{\rm PAM}({\bm k};{\bm p})
=\epsilon_{\alpha_1}^{\rm PAM}({\bm k}-{\bm p})+\epsilon_{\alpha_2}^{\rm PAM}({\bm p}), 
\end{align}
by using the electronic-mode energies defined in Eqs. (\ref{eq:eleEneKLM}) and (\ref{eq:eleEnePAM}) 
for the KLM and PAM, respectively. 
\par
The light-yellow region in Fig. \ref{fig:weff}(a) indicates the spin excitation continuum 
described by $\omega=e_{\rm ph}^{\rm KLM}({\bm k};{\bm p})$ for $J_{\rm K}/t=6$ in the 1D KLM. 
The validity is confirmed using the non-Abelian DDMRG method [Fig. \ref{fig:weff}(b)]. 
\par
The continuum slightly above the spin mode in Fig. \ref{fig:Akw}(d) and that in Fig. \ref{fig:Akw}(h) 
can be primarily identified as the particle-hole excited states 
$|3_{\sigma}1_{{\sigma}^{\prime}}\rangle_{{\bm k};{\bm p}}^{\rm KLM}$ 
with $\omega=e_{\rm ph}^{\rm KLM}({\bm k};{\bm p})$ and 
$|3_{\sigma}^-1_{{\sigma}^{\prime}}^-\rangle_{{\bm k};{\bm p}}^{\rm PAM}$ 
with $\omega=e_{\rm ph}^{\rm PAM}({\bm k};{\bm p})$, respectively. 
\section{Doped Kondo insulator} 
\subsection{Ground state} 
The eigenstates in the low-$|\omega|$ electron-removal mode of the Kondo insulator ($n=1$) are effectively expressed as 
$|1_{\sigma}\rangle_{\bm k}^{\rm KLM}$ with $\omega=-\epsilon_{1_{\sigma}}^{\rm KLM}({\bm k})$ 
[Eqs. (\ref{eq:modeKLM}) and (\ref{eq:eleEneKLM})] for the KLM; 
$|1_{\sigma}^-\rangle_{\bm k}^{\rm PAM}$ with $\omega=-\epsilon_{1_{\sigma}^-}^{\rm PAM}({\bm k})$ 
[Eqs. (\ref{eq:modePAM}) and (\ref{eq:eleEnePAM})] for the PAM. 
Hence, the ground state of a hole-doped system can be effectively obtained by removing electrons 
from this mode above the Fermi level in the same way as that of a doped band insulator. 
\par
In a one-hole-doped system where an electron with spin $\sigma$ and momentum ${\bm k}_{\rm F}(={\bm \pi})$ is removed, 
the ground state is effectively obtained as 
$|1_{\bar \sigma}\rangle_{-{\bm k}_{\rm F}}^{\rm KLM}$ for the KLM; 
$|1_{\bar \sigma}^-\rangle_{-{\bm k}_{\rm F}}^{\rm PAM}$ for the PAM, 
where ${\bar \sigma}$ represents the spin opposite to $\sigma$. 
The chemical potential $\mu$ is adjusted such that 
$\epsilon_{1_{\bar \sigma}}^{\rm KLM}(-{\bm k}_{\rm F})=0$ for the KLM; 
$\epsilon_{1_{\bar \sigma}^-}^{\rm PAM}(-{\bm k}_{\rm F})=0$ for the PAM. 
Then, the ground-state energy of the one-hole-doped system is equal to that of the undoped system: 
$E_{{\rm GS},N_{\rm h}=1}^{\rm KLM}=E_{{\rm GS},N_{\rm h}=0}^{\rm KLM}$; 
$E_{{\rm GS},N_{\rm h}=1}^{\rm PAM}=E_{{\rm GS},N_{\rm h}=0}^{\rm PAM}$, 
where $E_{{\rm GS},N_{\rm h}=m}^{\rm KLM}$ and $E_{{\rm GS},N_{\rm h}=m}^{\rm PAM}$ denote 
the ground-state energies of the $m$-hole-doped system in the KLM and PAM, respectively. 
\subsection{Electronic excitation} 
The electron-removal and electron-addition excited states from the one-holed-doped ground state 
in the KLM can be obtained as 
\begin{align}
\label{eq:eleremKLM}
c_{{\bm k},\sigma^\prime}|1_{\bar \sigma}\rangle_{-{\bm k}_{\rm F}}^{\rm KLM}
&=\frac{1}{L}\sum_{i,j=1}^L\mathrm{e}^{-i{\bm k}\cdot{\bm r}_i}\mathrm{e}^{-i{{\bm k}_{\rm F}} \cdot{\bm r}_j}
c_{i,\sigma^\prime}|1_{\bar \sigma}\rangle_j\prod_{l\ne j}^{L}|{\rm S}\rangle_l\nonumber\\
&=\frac{1}{L}\sum_{i\ne j}^L
\mathrm{e}^{-i{\bm k}\cdot{\bm r}_i}\mathrm{e}^{-i{{\bm k}_{\rm F}}\cdot{\bm r}_j}
c_{i,\sigma^\prime}|{\rm S}\rangle_i|1_{\bar \sigma}\rangle_j\prod_{l\ne i,j}^{L}|{\rm S}\rangle_l,\\
\label{eq:eleaddKLM}
c_{{\bm k},\sigma^\prime}^\dagger|1_{\bar \sigma}\rangle_{-{\bm k}_{\rm F}}^{\rm KLM}
&=\frac{1}{L}\sum_{i,j=1}^L\mathrm{e}^{i{\bm k}\cdot{\bm r}_i}\mathrm{e}^{-i{\bm k}_{\rm F}\cdot{\bm r}_j}
c_{i,\sigma^\prime}^\dagger|1_{\bar \sigma}\rangle_j\prod_{l\ne j}^{L}|{\rm S}\rangle_l\nonumber\\
&=\frac{1}{L}\sum_{i\ne j}^L\mathrm{e}^{i{\bm k}\cdot{\bm r}_i}\mathrm{e}^{-i{\bm k}_{\rm F}\cdot{\bm r}_j}
c_{i,\sigma^\prime}^\dagger|{\rm S}\rangle_i|1_{\bar \sigma}\rangle_j\prod_{l\ne i,j}^{L}|{\rm S}\rangle_l\nonumber\\
&+\frac{1}{L}\sum_{i=1}^L\mathrm{e}^{i({\bm k}-{\bm k}_{\rm F})\cdot{\bm r}_i}
c_{i,\sigma^\prime}^\dagger|1_{\bar \sigma}\rangle_i\prod_{l\ne i}^{L}|{\rm S}\rangle_l.
\end{align}
\par
Because $c_{i,\sigma^\prime}|{\rm S}\rangle_i\propto|1_{\bar \sigma^\prime}\rangle_i$ (Table \ref{tbl:1site}) 
in Eq. (\ref{eq:eleremKLM}), the excitation energy of the electron-removal excitation 
from the one-hole-doped ground state $|1_{\bar \sigma}\rangle_{-{\bm k}_{\rm F}}^{\rm KLM}$ 
is approximately $\epsilon_{1_{\bar \sigma^\prime}}^{\rm KLM}(-{\bm k})$: 
the energy of Eq. (\ref{eq:eleremKLM}) 
[$\approx\epsilon_{1_{\bar \sigma^\prime}}^{\rm KLM}(-{\bm k})+\epsilon_{1_{\bar \sigma}}^{\rm KLM}(-{\bm k}_{\rm F})
+E_{{\rm GS},N_{\rm h}=0}^{\rm KLM}$ [Eq. (\ref{eq:EphKLM})]] minus the one-holed-doped ground-state energy 
$E_{{\rm GS},N_{\rm h}=1}^{\rm KLM}$, where $\epsilon_{1_{\bar \sigma}}^{\rm KLM}(-{\bm k}_{\rm F})=0$ and 
$E_{{\rm GS},N_{\rm h}=1}^{\rm KLM}=E_{{\rm GS},N_{\rm h}=0}^{\rm KLM}$. 
Similarly, the excitation energy of the electron-addition excitation described by the first term in Eq. (\ref{eq:eleaddKLM}) 
from the one-hole-doped ground state is approximately $\epsilon_{3_{\sigma^\prime}}^{\rm KLM}({\bm k})$ 
because $c_{i,\sigma^\prime}^\dagger|{\rm S}\rangle_i\propto|3_{\sigma^\prime}\rangle_i$ (Table \ref{tbl:1site}). 
Thus essentially the same electronic modes as those of the Kondo insulator 
[Eq. (\ref{eq:eleEneKLM}) with $\alpha=1_{\bar \sigma^\prime}$ and $3_{\sigma^\prime}$; Sec. \ref{sec:modesKLM}] exist 
for both $\omega<0$ [Eq. (\ref{eq:eleremKLM})] and $\omega>0$ [the first term of Eq. (\ref{eq:eleaddKLM})] 
in the one-hole-doped system. 
\par
The dispersion relations of the electronic modes 
$\omega=\epsilon_{3_{\sigma^\prime}}^{\rm KLM}({\bm k})$ and 
$\omega=-\epsilon_{1_{\bar \sigma^\prime}}^{\rm KLM}(-{\bm k})$ are shown 
by the brown curves in Fig. \ref{fig:weff}(e) for $J_{\rm K}/t=6$ in the 1D KLM. 
The validity is confirmed using the non-Abelian DDMRG method [Fig. \ref{fig:weff}(f)]. 
\par
Similarly, the electron-removal and electron-addition excited states of the PAM can be obtained as 
\begin{align}
\label{eq:eleremPAM}
a_{{\bm k},\sigma^\prime}|1_{\bar \sigma}^-\rangle_{-{\bm k}_{\rm F}}^{\rm PAM}
&=\frac{1}{L}\sum_{i\ne j}^L\mathrm{e}^{-i{\bm k}\cdot{\bm r}_i}\mathrm{e}^{-i{\bm k}_{\rm F}\cdot{\bm r}_j}
a_{i,\sigma^\prime}|\psi_U^-\rangle_i|1_{\bar \sigma}^-\rangle_j\prod_{l\ne i,j}^{L}|\psi_U^-\rangle_l\nonumber\\
&+\frac{1}{L}\sum_{i=1}^L\mathrm{e}^{i(-{\bm k}-{\bm k}_{\rm F})\cdot{\bm r}_i}
a_{i,\sigma^\prime}|1_{\bar \sigma}^-\rangle_i\prod_{l\ne i}^{L}|\psi_U^-\rangle_l,\\
a_{{\bm k},\sigma^\prime}^\dagger|1_{\bar \sigma}^-\rangle_{-{\bm k}_{\rm F}}^{\rm PAM}
&=\frac{1}{L}\sum_{i\ne j}^L\mathrm{e}^{i{\bm k}\cdot{\bm r}_i}\mathrm{e}^{-i{\bm k}_{\rm F}\cdot{\bm r}_j}
a_{i,\sigma^\prime}^\dagger|\psi_U^-\rangle_i|1_{\bar \sigma}^-\rangle_j\prod_{l\ne i,j}^{L}|\psi_U^-\rangle_l\nonumber\\
&+\frac{1}{L}\sum_{i=1}^L\mathrm{e}^{i({\bm k}-{\bm k}_{\rm F})\cdot{\bm r}_i}
a_{i,\sigma^\prime}^\dagger|1_{\bar \sigma}^-\rangle_i\prod_{l\ne i}^{L}|\psi_U^-\rangle_l,
\label{eq:eleaddPAM}
\end{align}
where $a$ represents $c$ and $f$. 
It should be noted that $a_{i,\sigma^\prime}|\psi_U^-\rangle_i$ can be expressed as a linear combination of 
$|1_{\bar \sigma^\prime}^+\rangle_i$ and $|1_{\bar \sigma^\prime}^-\rangle_i$ for $U=2\Delta$ and $\infty$ 
and that $a_{i,\sigma^\prime}^\dagger|\psi_U^-\rangle_i$ can be expressed as a linear combination of 
$|3_{\sigma^\prime}^+\rangle_i$ and $|3_{\sigma^\prime}^-\rangle_i$ for $U=2\Delta$; 
$a_{i,\sigma^\prime}^\dagger|\psi_U^-\rangle_i\propto |3_{\sigma^\prime}\rangle_i$ for $U=\infty$ (Table \ref{tbl:1site}). 
\par
In the same way as the KLM, the excitation energies of the $|1_{\bar \sigma^\prime}^\pm\rangle$ components 
in the first term of Eq. (\ref{eq:eleremPAM}) are approximately 
$\epsilon_{1_{\bar \sigma^\prime}^\pm}^{\rm PAM}(-{\bm k})$ for $U=2\Delta$ and $\infty$; 
those of the $|3_{\sigma^\prime}^\pm\rangle$ components in the first term of Eq. (\ref{eq:eleaddPAM}) 
are approximately $\epsilon_{3_{\sigma^\prime}^\pm}^{\rm PAM}({\bm k})$ for $U=2\Delta$; 
that of the first term in Eq. (\ref{eq:eleaddPAM}) is approximately 
$\epsilon_{3_{\sigma^\prime}}^{\rm PAM}({\bm k})$ for $U=\infty$. 
\par
Thus, in both KLM and PAM, essentially the same electronic modes as those of Kondo insulators 
(Secs. \ref{sec:modesKLM} and \ref{sec:modesPAM}) exist in the one-hole-doped systems, 
similar to the electronic modes in a doped band insulator. 
The above argument can be extended to the $m$-hole-doped systems for $m\ll L$. 
\par
In Fig. \ref{fig:Akw}(b), the mode for $\omega<0$ corresponds to Eq. (\ref{eq:eleremKLM}), 
and the upper mode for $\omega>0$ corresponds to the first term of Eq. (\ref{eq:eleaddKLM}). 
In Fig. \ref{fig:Akw}(f), the low-$|\omega|$ and highest-$|\omega|$ modes for $\omega<0$ correspond to 
the $|1_{\bar \sigma^\prime}^-\rangle$ and $|1_{\bar \sigma^\prime}^+\rangle$ components 
in the first term of Eq. (\ref{eq:eleremPAM}), respectively. 
The dominant mode for $2\lesssim\omega\lesssim 4$ and the highest-$\omega$ mode for $\omega>0$ correspond to 
the $|3_{\sigma^\prime}^-\rangle$ and $|3_{\sigma^\prime}^+\rangle$ components 
in the first term of Eq. (\ref{eq:eleaddPAM}), respectively. 
\subsection{Doping-induced modes in the KLM} 
\label{sec:DISKLM}
The results in Figs. \ref{fig:Akw}(b) and \ref{fig:Akw}(f) contain emergent modes that cannot be identified 
as essentially the same modes as those of Kondo insulators. 
The doping-induced modes can be explained by the second terms of 
Eqs. (\ref{eq:eleaddKLM})--(\ref{eq:eleaddPAM}). 
\par
In the KLM, because 
\begin{align}
\label{eq:addcKLM}
c_{i,\sigma^\prime}^\dagger|1_{\bar \sigma}\rangle_i=&|\sigma^\prime,{\bar \sigma}\rangle_i\nonumber\\
=&\left\{
\begin{array}{lcc}
|{\rm T}_1\rangle&{\rm for }&(\sigma^\prime,{\bar \sigma})=(\uparrow,\uparrow)\\
|{\rm T}_{-1}\rangle&{\rm for }&(\sigma^\prime,{\bar \sigma})=(\downarrow,\downarrow)\\
\frac{1}{\sqrt{2}}(|{\rm T}_0\rangle+|{\rm S}\rangle)&{\rm for }&(\sigma^\prime,{\bar \sigma})=(\uparrow,\downarrow)\\
\frac{1}{\sqrt{2}}(|{\rm T}_0\rangle-|{\rm S}\rangle)&{\rm for }&(\sigma^\prime,{\bar \sigma})=(\downarrow,\uparrow)
\end{array}\right.
\end{align}
(Table \ref{tbl:1site}), 
when an electron with spin $\sigma^\prime$ parallel to that of the one-hole-doped ground state ($\bar \sigma$) is added, 
the second term of Eq. (\ref{eq:eleaddKLM}) is proportional to 
the spin excited state of the Kondo insulator [Eq. (\ref{eq:modeKLM}) 
with $\alpha={\rm T}_1$ or ${\rm T}_{-1}$ and momentum ${\bm k}-{\bm k}_{\rm F}$]. 
When an electron with spin $\sigma^\prime$ opposite to that of the one-hole-doped ground state is added, 
the second term of Eq. (\ref{eq:eleaddKLM}) has a component proportional to 
the spin excited state of the Kondo insulator [Eq. (\ref{eq:modeKLM}) with $\alpha={\rm T}_0$ 
and momentum ${\bm k}-{\bm k}_{\rm F}$] 
and a component proportional to the ground state of the Kondo insulator [Eq. (\ref{eq:GSKLM})] if ${\bm k}={\bm k}_{\rm F}$. 
\par
This implies that the spin excited states of the Kondo insulator appear in the electron-addition spectrum 
of the one-hole-doped system and exhibit the dispersion relation of the spin excitation [Eq. (\ref{eq:spinEneKLM})] 
but shifted by the Fermi momentum ${\bm k}_{\rm F}$: 
\begin{equation}
\label{eq:disSpinDISKLM}
\omega=e_{\rm spin}^{\rm KLM}({\bm k}-{\bm k}_{\rm F}).
\end{equation}
This characteristic is the same as that of the Mott transition 
\cite{Kohno1DHub,Kohno2DHub,KohnoDIS,KohnoAF,KohnoRPP,KohnoHubLadder,KohnoSpin,Kohno2DtJ,Kohno1DtJ,KohnoGW}. 
\par
The dispersion relation of the doping-induced mode 
$\omega=e_{\rm spin}^{\rm KLM}({\bm k}-{\bm k}_{\rm F})$ is shown 
by the blue curve in Fig. \ref{fig:weff}(e) for $J_{\rm K}/t=6$ in the 1D KLM. 
The validity is confirmed using the non-Abelian DDMRG method [Fig. \ref{fig:weff}(f)]. 
\subsection{Doping-induced modes in the PAM} 
\label{sec:DISPAM}
In the PAM, doping-induced electron-addition excited states of the one-hole-doped Kondo insulator can be explained 
by the second terms of Eq. (\ref{eq:eleaddPAM}). In this term, 
$c_{i,\sigma^\prime}^\dagger|1_{\bar \sigma}^-\rangle_i$ and $f_{i,\sigma^\prime}^\dagger|1_{\bar \sigma}^-\rangle_i$ 
have spin-triplet components ($|{\rm T}_1\rangle$, $|{\rm T}_{-1}\rangle$, or $|{\rm T}_0\rangle$) 
because 
\begin{align}
\label{eq:addcPAM}
c_{i,\sigma^\prime}^\dagger|1_{\bar \sigma}^-\rangle_i&=u_{t_{\rm K}}^+|\sigma^\prime,{\bar \sigma}\rangle_i
+2s_{\sigma^\prime}\delta_{\sigma^\prime,\sigma}u_{t_{\rm K}}^-|\uparrow\downarrow,0\rangle_i,\\
\label{eq:adddPAM}
f_{i,\sigma^\prime}^\dagger|1_{\bar \sigma}^-\rangle_i&=-u_{t_{\rm K}}^-|{\bar \sigma},\sigma^\prime\rangle_i\nonumber\\
&+2s_{\sigma^\prime}\delta_{\sigma^\prime,\sigma}(1-\delta_{U,\infty})u_{t_{\rm K}}^+|0,\uparrow\downarrow\rangle_i
\end{align}
for $U=2\Delta$ and $\infty$ (Table \ref{tbl:1site}). 
This implies that the spin excited states of the Kondo insulator appear in the electron-addition spectrum 
of the one-hole-doped system and exhibit the dispersion relation of the spin excitation [Eq. (\ref{eq:spinEnePAM})] 
but shifted by the Fermi momentum ${\bm k}_{\rm F}$: 
\begin{equation}
\label{eq:disSpinDISPAM}
\omega=e_{\rm spin}^{\rm PAM}({\bm k}-{\bm k}_{\rm F})
\end{equation}
in the PAM, as in the case of the KLM (Sec. \ref{sec:DISKLM}) and the Mott transition 
\cite{Kohno1DHub,Kohno2DHub,KohnoDIS,KohnoAF,KohnoRPP,KohnoHubLadder,KohnoSpin,Kohno2DtJ,Kohno1DtJ,KohnoGW}. 
\par
Equations (\ref{eq:addcPAM}) and (\ref{eq:adddPAM}) indicate that 
$c_{i,\sigma^\prime}^\dagger|1_{\bar \sigma}^-\rangle_i$ and $f_{i,\sigma^\prime}^\dagger|1_{\bar \sigma}^-\rangle_i$ 
also have the $|\psi_U^+\rangle$ and $|{\rm D}^-\rangle$ components for $U=2\Delta$ and 
the $|\psi_U^+\rangle$ component for $U=\infty$ when the spin of the added electron is opposite to that of the ground state. 
This implies that additional electron-addition modes can emerge upon doping, whose dispersion relation is expressed as 
\begin{equation}
\label{eq:disAlphaDISPAM}
\omega=e_{\alpha}^{\rm PAM}({\bm k}-{\bm k}_{\rm F}) 
\end{equation}
for $\alpha=\psi_U^+$ and ${\rm D}^-$. 
Here, $e_{\alpha}^{\rm PAM}({\bm k})$ is defined similarly to Eq. (\ref{eq:spinEnePAM}) as 
\begin{equation}
\label{eq:EneAlphaPAM}
e_{\alpha}^{\rm PAM}({\bm k})=I_{{\rm eff},\alpha}^{\rm PAM}d\gamma({\bm k})+{\bar E}_{\alpha}-{\bar E}_{\psi_U^-}
+2d(\xi_{\alpha\psi_U^-}-\xi_{\psi_U^-\psi_U^-}), 
\end{equation}
where $I_{{\rm eff},\alpha}^{\rm PAM}$, $\xi_{\alpha\psi_U^-}$, and $\xi_{\psi_U^-\psi_U^-}$ are 
quantities of $\mathcal{O}(t^2/\Delta{\bar E})$ obtained in the second-order perturbation theory. 
\par
It should be noted that the mode of $|\psi_{2\Delta}^+\rangle$ in the second term of Eq. (\ref{eq:eleaddPAM}) 
should be regarded as a part of the mode of $|3_{\sigma^\prime}^+\rangle$ in the first term of Eq. (\ref{eq:eleaddPAM}) 
at $U=2\Delta$ because the excitation energy of $|\psi_{2\Delta}^+\rangle$ from $|1_{\bar \sigma}^-\rangle$ in the second term 
is the same as that of $|3_{\sigma^\prime}^+\rangle$ from $|\psi_{2\Delta}^-\rangle$ in the first term (Table \ref{tbl:1site}): 
\begin{align}
{\bar E}_{\psi_{2\Delta}^+}-{\bar E}_{1_{\bar \sigma}^-}&=E_{2t_{\rm K}}^+-E_{t_{\rm K}}^--\mu\nonumber\\
&=E_{t_{\rm K}}^+-E_{2t_{\rm K}}^--\mu={\bar E}_{3_{\sigma^\prime}^+}-{\bar E}_{\psi_{2\Delta}^-}.
\end{align}
\par
Thus, in the high-$\omega$ regime for $\omega>0$, 
the $|{\rm D}^-\rangle_{{\bm k}-{\bm k}_F}^{\rm PAM}$ mode [Eq. (\ref{eq:modePAM}) with $\alpha={\rm D}^-$] 
exhibiting the dispersion relation in Eq. (\ref{eq:disAlphaDISPAM}) with $\alpha={\rm D}^-$ is 
induced by doping for $U=2\Delta$; 
the $|\psi_{\infty}^+\rangle_{{\bm k}-{\bm k}_F}^{\rm PAM}$ mode [Eq. (\ref{eq:modePAM}) with $\alpha=\psi_{\infty}^+$] 
exhibiting the dispersion relation in Eq. (\ref{eq:disAlphaDISPAM}) with $\alpha=\psi_{\infty}^+$ is 
induced by doping for $U=\infty$. 
\par
In the PAM, an electronic mode is induced by doping not only in the electron-addition spectrum ($\omega>0$) 
but also in the electron-removal spectrum ($\omega<0$). 
The $|0\rangle_{-{\bm k}-{\bm k}_F}^{\rm PAM}$ mode [Eq. (\ref{eq:modePAM}) with $\alpha=0$] 
is induced by doping for $\omega<0$, exhibiting the dispersion relation 
\begin{equation}
\label{eq:disAlpharemDISPAM}
\omega=-e_{0}^{\rm PAM}(-{\bm k}-{\bm k}_{\rm F}) 
\end{equation}
because
\begin{align}
c_{i,\sigma^\prime}|1_{\bar \sigma}^-\rangle_i&=\delta_{\sigma^\prime,{\bar \sigma}}u_{t_{\rm K}}^-|0,0\rangle_i,\\
f_{i,\sigma^\prime}|1_{\bar \sigma}^-\rangle_i&=\delta_{\sigma^\prime,{\bar \sigma}}u_{t_{\rm K}}^+|0,0\rangle_i
\end{align}
in the second term of Eq. (\ref{eq:eleremPAM}) (Table \ref{tbl:1site}). 
\subsection{Identification of doping-induced modes} 
From the viewpoint of small $t$, the doping-induced modes in Figs. \ref{fig:Akw}(b) and \ref{fig:Akw}(f) 
can be identified as follows. 
\par
In the low-$|\omega|$ regime for $\omega>0$, 
the doping-induced in-gap mode in Fig. \ref{fig:Akw}(b) [Fig. \ref{fig:Akw}(f)] corresponds to 
the mode of the spin-triplet state ($|{\rm T}_1\rangle$, $|{\rm T}_{-1}\rangle$, or $|{\rm T}_0\rangle$) 
in the second term of Eq. (\ref{eq:eleaddKLM}) for the KLM [Eq. (\ref{eq:eleaddPAM}) for PAM]. 
This mode in the KLM as well as in the PAM originates from the spin excited states of the Kondo insulator 
[Eqs. (\ref{eq:modeKLM}) and (\ref{eq:modePAM}) with $\alpha={\rm T}_1$, ${\rm T}_{-1}$, or ${\rm T}_0$] 
and essentially exhibits the spin-mode dispersion relation [Eqs. (\ref{eq:spinEneKLM}) and (\ref{eq:spinEnePAM})] 
but shifted by the Fermi momentum ${\bm k}_{\rm F}$ [Eqs. (\ref{eq:disSpinDISKLM}) and (\ref{eq:disSpinDISPAM})]. 
\par
In the PAM, additional electronic modes are induced by doping in the high-$|\omega|$ regime. 
The high-$|\omega|$ doping-induced modes in Fig. \ref{fig:Akw}(f) correspond 
to the $|{\rm D}^-\rangle_{{\bm k}-{\bm k}_F}^{\rm PAM}$ mode [the second term of Eq. (\ref{eq:eleaddPAM})] 
exhibiting the dispersion relation in Eq. (\ref{eq:disAlphaDISPAM}) with $\alpha={\rm D}^-$ for $\omega>0$ 
and the $|0\rangle_{-{\bm k}-{\bm k}_F}^{\rm PAM}$ mode [the second term of Eq. (\ref{eq:eleremPAM})] 
exhibiting the dispersion relation in Eq. (\ref{eq:disAlpharemDISPAM}) for $\omega<0$. 
\section{Energy scales and spectral weight} 
\label{sec:eneScalesSW}
\subsection{Energy scales in Kondo insulators} 
\label{sec:eneScalesIns}
\begin{figure}
\begin{tabular}{c|lll}
$\omega$&Electron ($n=1$)&Spin ($n=1$)&Electron ($n\rightarrow 1$)\\
$\uparrow$&$|3_{\sigma}^+\rangle$&&$|3_{\sigma}^+\rangle$\\
&&&$\updownarrow{\cal E}_{\rm spin}^{U=2\Delta}$\\
&$\updownarrow\sqrt{\Delta^2+4t_{\rm K}^2}$&&$|{\rm D}^-\rangle$\\\\
&$|3_{\sigma}^-\rangle$&&$|3_{\sigma}^-\rangle$\\\\
&$\updownarrow{\cal E}_{\rm charge}^{U=2\Delta}$&$|{\rm T}_{2s_\sigma}\rangle$&$|{\rm T}_{2s_\sigma}\rangle$\\
&&$\updownarrow{\cal E}_{\rm spin}^{U=2\Delta}$&$\updownarrow{\cal E}_{\rm spin}^{U=2\Delta}$\\
0&$|1_{\sigma}^-\rangle$&$|\psi_{2\Delta}^-\rangle$&$|1_{\sigma}^-\rangle$\\\\
&$\updownarrow\sqrt{\Delta^2+4t_{\rm K}^2}$&&$|0\rangle$\\
&&&$\updownarrow{\cal E}_{\rm spin}^{U=2\Delta}$\\
&$|1_{\sigma}^+\rangle$&&$|1_{\sigma}^+\rangle$\\\hline
\end{tabular}
\footnote[0]{
\begin{tabular}{l}
${\cal E}_{\rm spin}^{U=2\Delta}=\frac{\sqrt{\Delta^2+16t_{\rm K}^2}-\Delta}{2}
\overset{t_{\rm K}\ll\Delta}{\approx}\frac{4t_{\rm K}^2}{\Delta}=J_{{\rm K}{\rm eff}}^{U=2\Delta}.$\\
${\cal E}_{\rm charge}^{U=2\Delta}=\sqrt{\Delta^2+16t_{\rm K}^2}-\sqrt{\Delta^2+4t_{\rm K}^2}
\overset{t_{\rm K}\ll\Delta}{\approx}\frac{3}{2}J_{{\rm K}{\rm eff}}^{U=2\Delta}.$\\
$\mu=-\frac{1}{2}{\cal E}_{\rm charge}^{U=2\Delta}.$
\end{tabular}}
\caption{Energy scales of electronic and spin modes in the small-$t$ limit for $U=2\Delta$ in the PAM 
represented by eigenstates on a site (Table \ref{tbl:1site}).}
\label{fig:energyscales}
\end{figure}
The energy scales of the characteristic modes in the small-$t$ regime 
can be estimated by using the energies on a site (Table \ref{tbl:1site}). 
The energy scales of the modes at half filling in the KLM and PAM can be estimated as 
\begin{align}
\label{eq:dEKLM}
\Delta E_{\alpha}^{\rm KLM}&={\bar E_\alpha}-{\bar E_{\rm S}},\\
\label{eq:dEPAM}
\Delta E_{\alpha}^U&={\bar E_\alpha}-{\bar E_{\psi_U^-}}, 
\end{align}
respectively, where $\alpha$ denotes an excited state on a site. 
\par
By putting $\alpha={\rm T}_+$, ${\rm T}_+$, or ${\rm T}_0$ in Eqs. (\ref{eq:dEKLM}) and (\ref{eq:dEPAM}), 
the energy scales of the spin excitation can be estimated as 
\begin{align}
\label{eq:effJKLM}
{\cal E}_{\rm spin}^{\rm KLM}&=J_{\rm K},\\
\label{eq:effJU2D}
{\cal E}_{\rm spin}^{U=2\Delta}&=\frac{\sqrt{\Delta^2+16t_{\rm K}^2}-\Delta}{2}
\overset{t_{\rm K}\ll\Delta}{\approx}\frac{4t_{\rm K}^2}{\Delta}=J_{{\rm K}{\rm eff}}^{U=2\Delta},\\
\label{eq:effJUinf}
{\cal E}_{\rm spin}^{U=\infty}&=\frac{\sqrt{\Delta^2+8t_{\rm K}^2}-\Delta}{2}
\overset{t_{\rm K}\ll\Delta}{\approx}\frac{2t_{\rm K}^2}{\Delta}=J_{{\rm K}{\rm eff}}^{U=\infty}
\end{align}
for the KLM, $U=2\Delta$ PAM, and $U=\infty$ PAM, respectively (Fig. \ref{fig:energyscales}). 
These energies correspond to the zeroth-order terms of perturbation theory 
in Eqs. (\ref{eq:spinEneKLM}) and (\ref{eq:spinEnePAM}). 
In the $t_{\rm K}\ll \Delta$ regime, the effective $J_{\rm K}$ of the PAM is 
$J_{{\rm K}{\rm eff}}^{U=2\Delta}=\frac{4t_{\rm K}^2}{\Delta}$ for $U=2\Delta$ [Eq. (\ref{eq:effJU2D})] \cite{TsunetsuguRMP}; 
$J_{{\rm K}{\rm eff}}^{U=\infty}=\frac{2t_{\rm K}^2}{\Delta}$ for $U=\infty$ [Eq. (\ref{eq:effJUinf})] 
if $t\ll\frac{t_{\rm K}^2}{\Delta}$. 
\par
By putting $\alpha=1_\sigma^{(-)}$ and $3_\sigma^{(-)}$ in Eqs. (\ref{eq:dEKLM}) and (\ref{eq:dEPAM}), 
the energy scales of the charge (electronic) excitation gap can be estimated as 
\begin{align}
\label{eq:chargeEne1siteKLM}
{\cal E}_{\rm charge}^{\rm KLM}&=\Delta E_{1_\sigma}^{\rm KLM}+\Delta E_{3_\sigma}^{\rm KLM}=\frac{3}{2}J_{\rm K},\\
\label{eq:chargeEne1sitePAMU2D}
{\cal E}_{\rm charge}^{U=2\Delta}
&=\Delta E_{1_\sigma^-}^{U=2\Delta}+\Delta E_{3_\sigma^-}^{U=2\Delta}\nonumber\\
&=\sqrt{\Delta^2+16t_{\rm K}^2}-\sqrt{\Delta^2+4t_{\rm K}^2}
\overset{t_{\rm K}\ll\Delta}{\approx}\frac{3}{2}J_{{\rm K}{\rm eff}}^{U=2\Delta},\\
\label{eq:chargeEne1sitePAMUinf}
{\cal E}_{\rm charge}^{U=\infty}
&=\Delta E_{1_\sigma^-}^{U=\infty}+\Delta E_{3_\sigma}^{U=\infty}\nonumber\\
&=\sqrt{\Delta^2+8t_{\rm K}^2}-\frac{\sqrt{\Delta^2+4t_{\rm K}^2}+\Delta}{2}
\overset{t_{\rm K}\ll\Delta}{\approx}\frac{3}{2}J_{{\rm K}{\rm eff}}^{U=\infty}
\end{align}
for the KLM, $U=2\Delta$ PAM, and $U=\infty$ PAM, respectively (Fig. \ref{fig:energyscales}). 
In the $t_{\rm K}\ll \Delta$ regime, the charge gap is approximately three halves of the effective $J_{\rm K}$ in the PAM 
[Eqs. (\ref{eq:chargeEne1sitePAMU2D}) and (\ref{eq:chargeEne1sitePAMUinf})] if $t\ll\frac{t_{\rm K}^2}{\Delta}$. 
\par
In the KLM, the charge (electronic) excitation gap 
$\epsilon_{3_{\sigma}}^{\rm KLM}({\bm 0})+\epsilon_{1_{\sigma}}^{\rm KLM}({\bm \pi})$ [Eq. (\ref{eq:eleEneKLM})] 
is known to be larger than the spin excitation energy $e_{\rm spin}^{\rm KLM}({\bm k})$ [Eq. (\ref{eq:spinEneKLM})] 
by 50\% in the limit of $t\ll J_{\rm K}$ \cite{TsunetsuguRMP}, 
which can be understood by considering the excitation energies on a single site 
[Eqs. (\ref{eq:effJKLM}) and (\ref{eq:chargeEne1siteKLM})]. 
\par
Similarly, in the PAM, because the charge excitation gap on a site 
[Eqs. (\ref{eq:chargeEne1sitePAMU2D}) and (\ref{eq:chargeEne1sitePAMUinf})] 
is approximately three halves of the spin excitation energy on a site [Eqs. (\ref{eq:effJU2D}) and (\ref{eq:effJUinf})], 
the electronic excitation gap 
[$\epsilon_{3_{\sigma}^-}^{\rm PAM}({\bm 0})+\epsilon_{1_{\sigma}^-}^{\rm PAM}({\bm \pi})$ for $U=2\Delta$; 
$\epsilon_{3_{\sigma}}^{\rm PAM}({\bm 0})+\epsilon_{1_{\sigma}^-}^{\rm PAM}({\bm \pi})$ for $U=\infty$] 
[Eq. (\ref{eq:eleEnePAM})] is larger than 
the spin excitation energy $e_{\rm spin}^{\rm PAM}({\bm k})$ [Eq. (\ref{eq:spinEnePAM})] 
for $t\ll t_{\rm K}\ll \Delta$ and $t\ll \frac{t_{\rm K}^2}{\Delta}$. 
\par
In the PAM at half filling, there are high-$|\omega|$ modes, whose energy scales can be estimated as 
\begin{align}
\label{eq:highOmega3Ene1sitePAM}
&\Delta E_{3_\sigma^+}^{U=2\Delta}+\mu=\Delta E_{1_\sigma^+}^{U=2\Delta}-\mu\nonumber\\
&=\frac{\sqrt{\Delta^2+4t_{\rm K}^2}+\sqrt{\Delta^2+16t_{\rm K}^2}}{2}\overset{t_{\rm K}\ll\Delta}{\approx}\Delta+\frac{5t_{\rm K}^2}{\Delta},\\
\label{eq:highOmega1Ene1sitePAM}
&\Delta E_{1_\sigma^+}^{U=\infty}-\mu\nonumber\\
&=\frac{\sqrt{\Delta^2+4t_{\rm K}^2}+\sqrt{\Delta^2+8t_{\rm K}^2}}{2}\overset{t_{\rm K}\ll\Delta}{\approx}\Delta+\frac{3t_{\rm K}^2}{\Delta}
\end{align}
for $U=2\Delta$ and $\infty$, respectively (Fig. \ref{fig:energyscales}). 
These modes are far from the Fermi level for $t\ll t_{\rm K}\ll \Delta$ 
[Eqs. (\ref{eq:highOmega3Ene1sitePAM}) and (\ref{eq:highOmega1Ene1sitePAM})]. 
\subsection{Energy scales in doped Kondo insulators} 
The energy scales of the doping-induced modes for the KLM and PAM (Secs. \ref{sec:DISKLM} and \ref{sec:DISPAM}) 
in the small-$t$ regime can be estimated as 
\begin{align}
\Delta {\tilde E}_\alpha^{\rm KLM}&={\bar E_\alpha}-{\bar E_{1_{\bar \sigma}}},\\
\Delta {\tilde E}_\alpha^U&={\bar E_\alpha}-{\bar E_{1_{\bar \sigma}^-}}, 
\end{align}
respectively, where $\alpha$ denotes an excited state on a site. 
Here, the chemical potential is adjusted such that 
${\bar E}_{1_{\bar \sigma}}={\bar E}_{\rm S}$ in the KLM; ${\bar E}_{1_{\bar \sigma}^-}={\bar E}_{\psi_U^-}$ in the PAM. 
\par
The doping-induced mode in the Kondo insulating gap originates from the spin excited states 
of the Kondo insulator. Hence, it has the same energy scale as the spin excitation 
[Eqs. (\ref{eq:effJKLM})--(\ref{eq:effJUinf})] (Fig. \ref{fig:energyscales}). 
\par
In the PAM, additional high-$|\omega|$ modes emerge on doping, whose energy scales can be estimated as 
\begin{align}
\label{eq:highOmegaEne1sitePAM}
&\Delta {\tilde E}_{{\rm D}^-}^{U=2\Delta}+\mu=\Delta {\tilde E}_{0}^{U=2\Delta}-\mu
=\Delta {\tilde E}_{0}^{U=\infty}-\mu\nonumber\\
&=\frac{\Delta+\sqrt{\Delta^2+4t_{\rm K}^2}}{2}\overset{t_{\rm K}\ll\Delta}{\approx}\Delta+\frac{t_{\rm K}^2}{\Delta}
\end{align}
(Fig. \ref{fig:energyscales}). 
These modes are far from the Fermi level for $t\ll t_{\rm K}\ll \Delta$ [Eq. (\ref{eq:highOmegaEne1sitePAM})] 
but closer than the high-$|\omega|$ modes originating from the Kondo insulator 
[Eqs. (\ref{eq:highOmega3Ene1sitePAM}) and (\ref{eq:highOmega1Ene1sitePAM})] 
by the energy scale of the spin excitation [Eqs. (\ref{eq:effJU2D}) and (\ref{eq:effJUinf})] (Fig. \ref{fig:energyscales}): 
\begin{align}
\label{eq:dEhighOmegap2D}
\Delta E_{3_\sigma^+}^{U=2\Delta}-\Delta {\tilde E}_{{\rm D}^-}^{U=2\Delta}&={\cal E}_{\rm spin}^{U=2\Delta}
\overset{t_{\rm K}\ll\Delta}{\approx}J_{{\rm K}{\rm eff}}^{U=2\Delta},\\
\label{eq:dEhighOmegan2D}
\Delta E_{1_\sigma^+}^{U=2\Delta}-\Delta {\tilde E}_{0}^{U=2\Delta}&={\cal E}_{\rm spin}^{U=2\Delta}
\overset{t_{\rm K}\ll\Delta}{\approx}J_{{\rm K}{\rm eff}}^{U=2\Delta},\\
\label{eq:dEhighOmeganInfty}
\Delta E_{1_\sigma^+}^{U=\infty}-\Delta {\tilde E}_{0}^{U=\infty}&={\cal E}_{\rm spin}^{U=\infty}
\overset{t_{\rm K}\ll\Delta}{\approx}J_{{\rm K}{\rm eff}}^{U=\infty}.
\end{align}
\subsection{Spectral weight of doping-induced states} 
\label{sec:swDIS}
The spectral weight of the doping-induced states can be estimated 
by using the eigenstates on a site (Table \ref{tbl:1site}). 
\par
Because the second term of Eq. (\ref{eq:eleaddKLM}) can be expressed as 
\begin{align}
\label{eq:addDISpmodeKLM}
&\frac{1}{L}\sum_{i=1}^L\mathrm{e}^{i({\bm k}-{\bm k}_{\rm F})\cdot{\bm r}_i}
c_{i,{\bar \sigma}}^\dagger|1_{\bar \sigma}\rangle_i\prod_{l\ne i}^{L}|{\rm S}\rangle_l
=\frac{1}{\sqrt{L}}|{\rm T}_{2s_{\bar \sigma}}\rangle_{{\bm k}-{\bm k}_{\rm F}}^{\rm KLM},\\
\label{eq:addDIS0modeKLM}
&\frac{1}{L}\sum_{i=1}^L\mathrm{e}^{i({\bm k}-{\bm k}_{\rm F})\cdot{\bm r}_i}
c_{i,\sigma}^\dagger|1_{\bar \sigma}\rangle_i\prod_{l\ne i}^{L}|{\rm S}\rangle_l\nonumber\\
&=\frac{1}{\sqrt{2L}}\left[|{\rm T}_0\rangle_{{\bm k}-{\bm k}_{\rm F}}^{\rm KLM}
+2s_\sigma|{\rm S}\rangle_{{\bm k}-{\bm k}_{\rm F}}^{\rm KLM}\right] 
\end{align} 
[Eqs. (\ref{eq:modeKLM}) and (\ref{eq:addcKLM})], 
the spectral weight of the doping-induced states of the one-hole-doped system [Eq. (\ref{eq:addAkw})] is obtained as 
$\frac{3}{4L}$ at each ${\bm k}$ in the KLM. 
\par
Similarly, in the PAM, the second term of Eq. (\ref{eq:eleaddPAM}) can be expressed as 
\begin{align}
&\frac{1}{L}\sum_{i=1}^L\mathrm{e}^{i({\bm k}-{\bm k}_{\rm F})\cdot{\bm r}_i}
c_{i,{\bar \sigma}}^\dagger|1_{\bar \sigma}^-\rangle_i\prod_{l\ne i}^{L}|\psi_U^-\rangle_l
=\frac{u_{t_{\rm K}}^+}{\sqrt{L}}|{\rm T}_{2s_{\bar \sigma}}\rangle_{{\bm k}-{\bm k}_{\rm F}}^{\rm PAM},\\
&\frac{1}{L}\sum_{i=1}^L\mathrm{e}^{i({\bm k}-{\bm k}_{\rm F})\cdot{\bm r}_i}
f_{i,{\bar \sigma}}^\dagger|1_{\bar \sigma}^-\rangle_i\prod_{l\ne i}^{L}|\psi_U^-\rangle_l
=-\frac{u_{t_{\rm K}}^-}{\sqrt{L}}|{\rm T}_{2s_{\bar \sigma}}\rangle_{{\bm k}-{\bm k}_{\rm F}}^{\rm PAM},\\
\label{eq:addcDISmodePAM}
&\frac{1}{L}\sum_{i=1}^L\mathrm{e}^{i({\bm k}-{\bm k}_{\rm F})\cdot{\bm r}_i}
c_{i,\sigma}^\dagger|1_{\bar \sigma}^-\rangle_i\prod_{l\ne i}^{L}|\psi_U^-\rangle_l\nonumber\\
&=\frac{u_{t_{\rm K}}^+}{\sqrt{2L}}\left[|{\rm T}_0\rangle_{{\bm k}-{\bm k}_{\rm F}}^{\rm PAM}
+2s_\sigma|{\rm S}\rangle_{{\bm k}-{\bm k}_{\rm F}}^{\rm PAM}\right]+\frac{2s_\sigma u_{t_{\rm K}}^-}{\sqrt{L}}|\uparrow\downarrow,0\rangle_{{\bm k}-{\bm k}_{\rm F}}^{\rm PAM},\\
\label{eq:adddDISmodePAM}
&\frac{1}{L}\sum_{i=1}^L\mathrm{e}^{i({\bm k}-{\bm k}_{\rm F})\cdot{\bm r}_i}
f_{i,\sigma}^\dagger|1_{\bar \sigma}^-\rangle_i\prod_{l\ne i}^{L}|\psi_U^-\rangle_l\nonumber\\
&=-\frac{u_{t_{\rm K}}^-}{\sqrt{2L}}\left[|{\rm T}_0\rangle_{{\bm k}-{\bm k}_{\rm F}}^{\rm PAM}
-2s_\sigma|{\rm S}\rangle_{{\bm k}-{\bm k}_{\rm F}}^{\rm PAM}\right]\nonumber\\
&+(1-\delta_{U,\infty})\frac{2s_\sigma u_{t_{\rm K}}^+}{\sqrt{L}}
|0,\uparrow\downarrow\rangle_{{\bm k}-{\bm k}_{\rm F}}^{\rm PAM}
\end{align}
for $U=2\Delta$ and $\infty$ [Eqs. (\ref{eq:modePAM}), (\ref{eq:addcPAM}), and (\ref{eq:adddPAM})]. 
Thus the spectral weight of the spin-triplet doping-induced states of the one-hole-doped system is 
obtained as $\frac{3}{4L}$ at each ${\bm k}$ in the PAM. 
\par
Furthermore, the spectral weight of the doping-induced $|{\rm D}^-\rangle_{{\bm k}-{\bm k}_F}^{\rm PAM}$ mode 
of the one-hole-doped system is obtained as $\frac{1}{4L}$ at each ${\bm k}$ for $U=2\Delta$ in the PAM because 
\begin{align}
|\uparrow\downarrow,0\rangle_{{\bm k}-{\bm k}_{\rm F}}^{\rm PAM}
&=\frac{1}{\sqrt{2}}\left[|{\rm D}^+\rangle_{{\bm k}-{\bm k}_{\rm F}}^{\rm PAM}
-|{\rm D}^-\rangle_{{\bm k}-{\bm k}_{\rm F}}^{\rm PAM}\right],\\
|0,\uparrow\downarrow\rangle_{{\bm k}-{\bm k}_{\rm F}}^{\rm PAM}
&=\frac{1}{\sqrt{2}}\left[|{\rm D}^+\rangle_{{\bm k}-{\bm k}_{\rm F}}^{\rm PAM}
+|{\rm D}^-\rangle_{{\bm k}-{\bm k}_{\rm F}}^{\rm PAM}\right].
\end{align}
For $U=\infty$, the spectral weight of the doping-induced 
$|\psi_\infty^+\rangle_{{\bm k}-{\bm k}_{\rm F}}^{\rm PAM}$ mode of the one-hole-doped system is 
$\mathcal{O}(\frac{t_{\rm K}^4}{L\Delta^4})$ for $t\ll t_{\rm K}\ll \Delta$. 
\par
In the $\omega<0$ regime, the spectral weight of the doping-induced mode of the one-hole-doped system is obtained as 
$\frac{1}{2L}$ at each ${\bm k}$ for $U=2\Delta$ and $\infty$ in the PAM 
because the second term of Eq. (\ref{eq:eleremPAM}) can be expressed as 
\begin{align}
&\frac{1}{L}\sum_{i=1}^L\mathrm{e}^{i(-{\bm k}-{\bm k}_{\rm F})\cdot{\bm r}_i}
c_{i,{\bar \sigma}}|1_{\bar \sigma}^-\rangle_i\prod_{l\ne i}^{L}|\psi_U^-\rangle_l
=\frac{u_{t_{\rm K}}^-}{\sqrt{L}}|0\rangle_{-{\bm k}-{\bm k}_{\rm F}}^{\rm PAM},\\
&\frac{1}{L}\sum_{i=1}^L\mathrm{e}^{i(-{\bm k}-{\bm k}_{\rm F})\cdot{\bm r}_i}
f_{i,{\bar \sigma}}|1_{\bar \sigma}^-\rangle_i\prod_{l\ne i}^{L}|\psi_U^-\rangle_l
=\frac{u_{t_{\rm K}}^+}{\sqrt{L}}|0\rangle_{-{\bm k}-{\bm k}_{\rm F}}^{\rm PAM}.
\end{align}
\subsection{Doping dependence of spectral weight} 
\label{sec:dopingDep}
The doping-induced states are obtained by adding an electron to or removing an electron from the doped site 
[$|1_{\bar \sigma}\rangle$ in Eq. (\ref{eq:eleaddKLM}) for the KLM; 
$|1_{\bar \sigma}^-\rangle$ in Eqs. (\ref{eq:eleremPAM}) and (\ref{eq:eleaddPAM}) for the PAM]. 
The probability of adding an electron to or removing an electron from the doped sites is 
proportional to the number of doped sites. 
Hence, the spectral weight of the doping-induced modes is expected to increase in proportion to the number of holes 
($N_{\rm h}$) in the small-doping regime ($\delta=\frac{N_{\rm h}}{2L}\ll 1$). 
In fact, the numerical results indicate that the spectral weight of the doping-induced modes increases as the doping 
concentration increases [Figs. \ref{fig:Akw}(a), \ref{fig:Akw}(b), \ref{fig:Akw}(e), \ref{fig:Akw}(f), and \ref{fig:weff}(f)]. 
\par
Based on the results obtained in Sec. \ref{sec:swDIS}, 
the spectral weight of the doping-induced modes at each ${\bm k}$ is expected to be 
$\frac{3\delta}{2}$ for the in-gap mode originating from the spin mode in the KLM and PAM; 
$\frac{\delta}{2}$ for the high-$\omega$ doping-induced mode in the $\omega>0$ regime 
($|{\rm D}^-\rangle_{{\bm k}-{\bm k}_F}^{\rm PAM}$ ) for $U=2\Delta$ in the PAM; 
$\delta$ for the high-$|\omega|$ doping-induced mode in the $\omega<0$ regime 
($|0\rangle_{-{\bm k}-{\bm k}_F}^{\rm PAM}$) for $U=2\Delta$ and $\infty$ in the PAM. 
\par
Because of the electronic anticommutation relations 
$\{c_{{\bm k},\sigma},c_{{\bm k},\sigma}^{\dagger}\}=1$ and 
$\{f_{{\bm k},\sigma},f_{{\bm k},\sigma}^{\dagger}\}=1$, 
the total spectral weight at each ${\bm k}$ follows the sum rule [Eqs. (\ref{eq:addAkw}) and (\ref{eq:remAkw})]: 
\begin{equation}
\label{eq:sumrule}
\int_{-\infty}^\infty d\omega A({\bm k},\omega)=
\left\{\begin{array}{llllll}
1&{\rm for}&{\rm the}&{\rm KLM},&\\
2&{\rm for}&{\rm the}&{\rm PAM}\quad(U<\infty).
\end{array}\right.
\end{equation}
Thus the spectral weight of the electronic excited states, except for the doping-induced states, should be 
$1-\frac{3\delta}{2}$ in the KLM and $2-3\delta$ for $U=2\Delta$ in the PAM at each ${\bm k}$. 
\section{Parameter dependence} 
\subsection{$J_{\rm K}$ and $t_{\rm K}$} 
In the KLM at half filling, the spin and charge excitation energies are primarily determined by $J_{\rm K}$ 
in the small-$t$ regime because ${\cal E}_{\rm spin}^{\rm KLM}=J_{\rm K}$ [Eq. (\ref{eq:effJKLM})] and 
${\cal E}_{\rm charge}^{\rm KLM}=\frac{3}{2}J_{\rm K}$ [Eq. (\ref{eq:chargeEne1siteKLM})] \cite{TsunetsuguRMP}. 
Thus the spin and charge (electronic) excitation energies are expected to become higher 
for larger $J_{\rm K}$. In fact, the numerical results obtained using the non-Abelian DDMRG method indicate that 
the spin and electron-addition excitations in the Kondo insulator at $J_{\rm K}=6t$ 
[Figs. \ref{fig:weff}(b) and \ref{fig:weff}(d)] have higher energies than those at $J_{\rm K}=2t$ 
[Figs. \ref{fig:Akw}(c) and \ref{fig:Akw}(d)]. 
This also implies that the energy of the doping-induced electronic mode becomes higher for larger $J_{\rm K}$ 
because the doping-induced mode exhibits the spin-mode dispersion relation shifted by the Fermi momentum 
[Eq. (\ref{eq:disSpinDISKLM}) and the blue curve in Fig. \ref{fig:weff}(e); Figs. \ref{fig:Akw}(b) and \ref{fig:weff}(f)]. 
\par
\begin{figure}
\includegraphics[width=\linewidth]{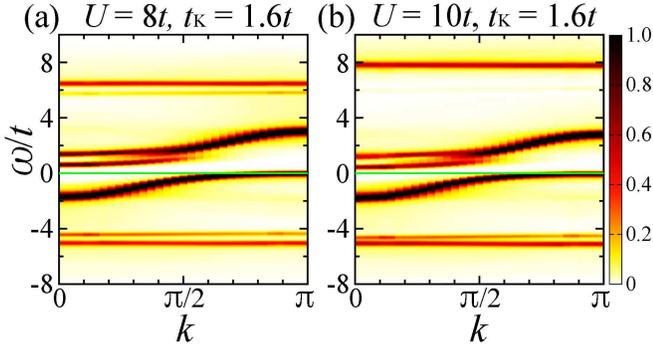}
\caption{$A(k,\omega)t$ of the 1D PAM at $n\approx 0.933$ 
for (a) $U=8t$, $t_{\rm K}=1.6t$, $\Delta=4t$ and (b) $U=10t$, $t_{\rm K}=1.6t$, $\Delta=4t$ 
obtained using the non-Abelian DDMRG method. The green lines indicate $\omega=0$ (Fermi level). 
Gaussian broadening with a standard deviation of $0.1t$ was used.}
\label{fig:tpUdep}
\end{figure}

Similarly, in the PAM in the small-$t$ regime, the energies of the low-$\omega$ modes for $\omega>0$ 
are expected to become higher for larger $J_{{\rm K}{\rm eff}}^U$ [Eqs. (\ref{eq:effJU2D}), (\ref{eq:effJUinf}), 
(\ref{eq:chargeEne1sitePAMU2D}), and (\ref{eq:chargeEne1sitePAMUinf})], 
which increases in proportion to $t_{\rm K}^2$ for large and fixed values of $U$ and $\Delta$ 
[Eqs. (\ref{eq:effJU2D}) and (\ref{eq:effJUinf})]. 
In fact, the numerical results obtained using the non-Abelian DDMRG method indicate that 
the low-$\omega$ doping-induced electronic mode for $\omega>0$, 
which essentially exhibits the spin-mode dispersion relation shifted by the Fermi momentum, 
and the low-$\omega$ electron-addition mode originating from the Kondo insulator for $\omega>0$ 
at $t_{\rm K}=2t$ [Fig. \ref{fig:Akw}(f)] have higher energies than those at $t_{\rm K}=1.6t$ [Fig. \ref{fig:tpUdep}(a)]. 
\par
In the PAM, because the high-$|\omega|$ doping-induced modes are closer to the Fermi level 
by the spin excitation energy [Eqs. (\ref{eq:dEhighOmegap2D})--(\ref{eq:dEhighOmeganInfty}); Fig. \ref{fig:energyscales}] 
than the high-$|\omega|$ modes originating from the Kondo insulator in the small-$t$ limit, 
their energy differences are expected to become larger as $t_{\rm K}$ increases 
for both $\omega>0$ and $\omega<0$ in the small-$t$ regime. 
In fact, the numerical results obtained using the non-Abelian DDMRG method indicate that 
the energy differences between the high-$|\omega|$ doping-induced mode and the high-$|\omega|$ mode originating from the Kondo insulator at $t_{\rm K}=2t$ [Fig. \ref{fig:Akw}(f)] are larger than those at $t_{\rm K}=1.6t$ [Fig. \ref{fig:tpUdep}(a)] 
for both $\omega>0$ and $\omega<0$. 
\subsection{$U\ne 2\Delta$} 
Next, we consider the interaction effect in the PAM. 
Because the $U$ term is related to double occupancy in the localized orbitals [Eq. (\ref{eq:PAM})], 
it would primarily affect the high-energy properties of $\mathcal{O}(U)$ related to the localized-orbital band 
exhibiting almost flat dispersion relation. 
As shown in Fig. \ref{fig:tpUdep}, the high-$\omega$ mode originating from the Kondo insulator 
for $\omega\approx 6.5t$ shifts to higher $\omega$, 
and the high-$\omega$ doping-induced mode for $\omega\approx 5.8t$ significantly loses spectral weight 
when $U$ is increased from $8t$ to $10t$. 
In the $\omega<0$ regime, the spectral features remain almost unchanged, 
but the energy difference between the high-$|\omega|$ modes around $\omega=-5t$ becomes smaller as $U$ increases. 
\par
The numerical results for $\omega>0$ in Fig. \ref{fig:tpUdep} and analyses in Sec. \ref{sec:eneScalesSW} indicate that 
the high-$\omega$ mode originating from $|3_\sigma^+\rangle_{{\bm k}}^{\rm PAM}$ 
with considerable spectral weight for $\omega>0$ shifts to higher $\omega$ as $U$ increases. 
Eventually, doubly occupied states in the localized orbitals 
(with infinitely high energy) are removed from the Hilbert space in the $U=\infty$ limit. 
This is why the electronic spectral-weight sum rule is not satisfied for $U=\infty$: 
$\int_{-\infty}^\infty d\omega A({\bm k},\omega)<2$. 
\par
The high-$\omega$ doping-induced mode originating from $|{\rm D}^-\rangle_{{\bm k}-{\bm k}_F}^{\rm PAM}$ 
for $\omega>0$ significantly loses spectral weight down to $\mathcal{O}\left(\frac{t_{\rm K}^4\delta}{\Delta^4}\right)$ 
for $U\rightarrow\infty$ (Secs. \ref{sec:swDIS} and \ref{sec:dopingDep}). 
As $\Delta$ increases, the doping-induced mode shifts to higher $\omega$, further losing spectral weight. 
Eventually, the spectral features are reduced to those of the KLM [Figs. \ref{fig:Akw}(b) and \ref{fig:weff}(f)] 
for $\Delta\rightarrow\infty$. 
\par
The numerical results for $\omega<0$ in Fig. \ref{fig:tpUdep} can be explained from the viewpoint of small $t$ as follows. 
Because the spectral weight of the high-$|\omega|$ modes for $\omega<0$ at $U=\infty$ is 
the same as that of $U=2\Delta$ in the small-$t$ limit (Sec. \ref{sec:swDIS}), 
the spectral weight would remain almost unchanged as $U$ increases in the small-$t$ regime. 
The energy difference between the high-$|\omega|$ modes ($|0\rangle_{-{\bm k}-{\bm k}_F}^{\rm PAM}$ and 
$|1_\sigma^+\rangle_{-{\bm k}}^{\rm PAM}$) at $U=2\Delta$ and that at $U=\infty$ are 
approximately $J_{{\rm K}{\rm eff}}^{U=2\Delta}$ and 
$J_{{\rm K}{\rm eff}}^{U=\infty}(=\frac{1}{2}J_{{\rm K}{\rm eff}}^{U=2\Delta})$ 
[Eqs. (\ref{eq:dEhighOmegan2D}) and (\ref{eq:dEhighOmeganInfty})], respectively, 
for $t_{\rm K}\ll\Delta$ and $t\ll\frac{t_{\rm K}^2}{\Delta}$. 
Hence, the energy difference would decrease as $U$ increases. 
The high-$|\omega|$ modes do not disappear even at $U=\infty$ (Secs. \ref{sec:swDIS} and \ref{sec:dopingDep}). 
As $\Delta$ increases, the high-$|\omega|$ modes shift to negatively larger $\omega$. 
Eventually, the modes with a vacant-localized-orbital state 
shift to $\omega\rightarrow-\infty$ and are removed from the Hilbert space for $\Delta\rightarrow\infty$; 
the spectral features are reduced to those of the KLM [Figs. \ref{fig:Akw}(b) and \ref{fig:weff}(f)]. 
\subsection{$U=0$} 
\label{sec:U0}
As $U$ decreases in the PAM, the spectral features should change into noninteracting bands. 
\begin{figure}
\includegraphics[width=\linewidth]{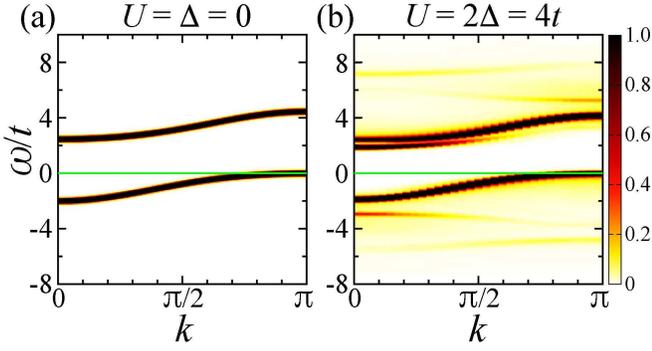}
\caption{$A(k,\omega)t$ of the 1D PAM at $n\approx 0.933$ 
for (a) $U=\Delta=0$, $t_{\rm K}=2t$ obtained using $\omega=\epsilon^\pm(k)$ [Eq. (\ref{eq:U0disp})] and 
(b) $U=2\Delta=4t$, $t_{\rm K}=2t$ obtained using the non-Abelian DDMRG method. 
The green lines indicate $\omega=0$ (Fermi level). 
Gaussian broadening with a standard deviation of $0.1t$ was used.}
\label{fig:Udep}
\end{figure}
In the noninteracting ($U=0$) PAM, the Hamiltonian [Eq. (\ref{eq:PAM})] can be diagonalized 
in the momentum space as \cite{TsunetsuguRMP} 
\begin{equation}
{\cal H}_{\rm PAM}\overset{U=0}{=}\sum_{{\bm k},\sigma,\alpha}
\epsilon^\alpha({\bm k})\beta_{{\bm k},\sigma}^{\alpha\dagger}\beta_{{\bm k},\sigma}^\alpha,
\end{equation}
where 
\begin{align}
\label{eq:U0disp}
\epsilon^\pm({\bm k})&=
-\frac{2td\gamma({\bm k})+\Delta}{2}\pm\frac{\sqrt{[-2td\gamma({\bm k})+\Delta]^2+4t_{\rm K}^2}}{2}-\mu,\\
\label{eq:U0beta}
\beta_{{\bm k},\sigma}^\pm&=u_{\bm k}^\pm c_{{\bm k},\sigma}\mp u_{\bm k}^\mp f_{{\bm k},\sigma},\\
u_{\bm k}^\pm&=\sqrt{\frac{1}{2}\left(1\pm\frac{-2td\gamma({\bm k})+\Delta}
{\sqrt{[-2td\gamma({\bm k})+\Delta]^2+4t_{\rm K}^2}}\right)}.
\end{align}
At half filling, the system becomes a band insulator for $t_{\rm K}\ne0$. 
The band structure does not change in the presence or absence of doping [Fig. \ref{fig:Udep}(a)]. 
\subsection{$U=2\Delta$} 
\label{sec:U2D}
By introducing the interaction term, the noninteracting mode for $\omega>0$ 
splits into two modes as shown in Fig. \ref{fig:Udep}(b). 
The lower-$\omega$ mode can be primarily identified 
as the doping-induced mode originating from the spin excitation of the Kondo insulator. 
In addition, the spectral weight emerges above (below) the noninteracting mode for $\omega>0$ ($\omega<0$); 
high-$|\omega|$ modes are formed around $\omega= 7t$ and $-5t$ 
and intermediate-$|\omega|$ modes around $\omega= 5t$ and $-3t$. 
By comparing Fig. \ref{fig:Udep}(b) with Fig. \ref{fig:Akw}(f), 
the high-$|\omega|$ modes can be primarily identified as the modes originating 
from $|3_\sigma^+\rangle_{\bm k}^{\rm PAM}$ and $|1_{\bar \sigma}^+\rangle_{-{\bm k}}^{\rm PAM}$ 
for $\omega>0$ and $\omega<0$, respectively. 
The intermediate-$|\omega|$ modes can be primarily identified as the modes originating 
from $|{\rm D}^-\rangle_{{\bm k}-{\bm k}_{\rm F}}^{\rm PAM}$ and $|0\rangle_{-{\bm k}-{\bm k}_{\rm F}}^{\rm PAM}$ 
for $\omega>0$ and $\omega<0$, respectively. 
\par
As $U$ and $\Delta$ increases with $U=2\Delta$, 
the high-$|\omega|$ and intermediate-$|\omega|$ modes shift to higher $|\omega|$. 
Eventually, these modes have infinitely high energies and are removed from the Hilbert space for $U=2\Delta\rightarrow\infty$; 
the spectral features are reduced to those of the KLM [Figs. \ref{fig:Akw}(b) and \ref{fig:weff}(f)]. 
\subsection{Conduction and localized orbitals} 
\begin{figure}
\includegraphics[width=\linewidth]{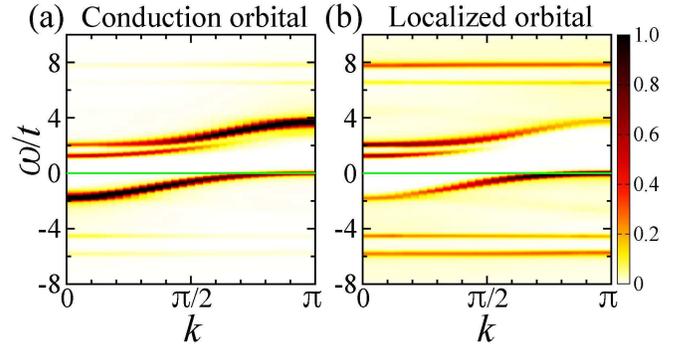}
\caption{Conduction-orbital and localized-orbital components of $A(k,\omega)t$ at $n\approx 0.933$ 
for $U=2\Delta=8t$, $t_{\rm K}=2t$ in the 1D PAM obtained using the non-Abelian DDMRG method. 
(a) Conduction-orbital component. (b) Localized-orbital component. 
The green lines indicate $\omega=0$ (Fermi level). 
Gaussian broadening with a standard deviation of $0.1t$ was used.}
\label{fig:cfcomp}
\end{figure}

The contributions from the conduction and localized orbitals are shown in Fig. \ref{fig:cfcomp}. 
The high-$|\omega|$ modes exhibiting almost flat dispersion relation are primarily due to the localized-orbital component 
for both $\omega>0$ and $\omega<0$ 
because these modes are mostly due to double occupancy and vacancy in the localized orbitals 
for $\omega>0$ and $\omega<0$, respectively. 
The conduction-orbital component is mainly located along the less-hybridized part of the dispersing modes. 
\par
Although a flat-dispersion part generally has a considerable localized-orbital component, 
the conduction-orbital component also contributes to the properties around the Fermi level, 
including the doping-induced mode originating from the spin excitation of the Kondo insulator. 
Even without the localized-orbital component in the large-$U$ and large-$\Delta$ limit, 
the characteristic spectral features in the low-$|\omega|$ regime are preserved 
in the conduction-orbital component, as shown in the KLM [Figs. \ref{fig:Akw}(b) and \ref{fig:weff}(f)]. 
\section{Discussion and summary} 
\subsection{Kondo-insulator--metal transition} 
By doping a Kondo insulator that has both spin and charge excitation gaps, 
holes are introduced around the top of the lower band, as in the case of a doped band insulator. 
Thus, as long as only ground-state properties are considered, 
the Kondo-insulator--metal transition is similar to the band-insulator--metal transition. 
Usually, the electronic excitation spectrum is also considered to remain essentially unchanged with or without doping, 
as in the case of a band insulator. 
However, as demonstrated in this paper, a remarkable change is introduced in the electronic excitation spectrum by doping: 
an electronic mode that exhibits momentum-shifted magnetic dispersion relation emerges in the Kondo insulating gap, 
reflecting the spin-charge separation of the Kondo insulator (different lowest excitation energies between spin and charge). 
This characteristic does not appear in a doped band insulator 
because the spin and charge excitation energies are the same as the band gap. 
Hence, this is a crucial strong-correlation effect by which the Kondo-insulator--metal transition can be distinguished 
from a simple band-insulator--metal transition. This characteristic is the same as that of the Mott transition 
\cite{Kohno1DHub,Kohno2DHub,KohnoDIS,KohnoAF,KohnoRPP,KohnoHubLadder,KohnoSpin,Kohno2DtJ,Kohno1DtJ,KohnoGW}. 
\par
Although the emergence of electronic states on doping a Mott insulator has been recognized since the early 1990s \cite{Eskes,DagottoDOS}, 
interpretations have been controversial \cite{Kohno1DHub,Kohno2DHub,KohnoDIS,KohnoAF,KohnoRPP,KohnoHubLadder,KohnoSpin,Kohno2DtJ,Kohno1DtJ,KohnoGW,SakaiImadaPRL,SakaiImadaPRB,ImadaCofermionPRL,ImadaCofermionPRB,PhillipsRMP,PhillipsRPP,EderOhta2DHub,EderOhtaIPES,2orbKanamoriHub}. 
The present results support the interpretation that the spin excitation whose energy is lower than the charge gap 
emerges as electronic excitation exhibiting momentum-shifted dispersion relation by doping 
\cite{Kohno1DHub,Kohno2DHub,KohnoDIS,KohnoAF,KohnoRPP,KohnoHubLadder,KohnoSpin,Kohno2DtJ,Kohno1DtJ,KohnoGW} 
and imply that this characteristic is general in doped spin-charge-separated insulators. 
\subsection{Effective mass} 
In general, as the interaction becomes stronger, the effective mass of the quasiparticle is considered to increase: 
the slope of the dispersion relation at the Fermi level becomes flatter, 
and the density of states increases \cite{LandauFL,Nozieres}. 
In particular, it is widely believed that strong correlations make the effective mass very large 
in heavy-fermion systems described by the KLM or PAM. 
\par
From the viewpoint of the KLM, the flattening of the dispersion relation at the Fermi level near a Kondo insulator 
can be regarded as a consequence of the spin interaction 
because it becomes flatter by introducing the $J_{\rm K}$ term, but it is flatter for weaker interaction $J_{\rm K}(>0)$ 
[Eq. (\ref{eq:eleEneKLM})]. 
This contrasts with the conventional view that stronger interaction makes the effective mass larger. 
\par
From the viewpoint of the PAM, the effect of interaction $U$ 
on the flattening of the dispersion relation at the Fermi level is 
much weaker than the hybridization effect between conduction and localized orbitals ($t_{\rm K}$) 
as shown in Figs. \ref{fig:Akw} and \ref{fig:Udep}(a) 
in the parameter regime investigated in this paper ($t\ll t_{\rm K}\ll\Delta<U$ and $t\ll J_{{\rm K} {\rm eff}}^U$ or $J_{\rm K}$). 
In particular, the effective hopping integral is $t/2$ up to $\mathcal{O}(t)$ in both the KLM [Eq. (\ref{eq:eleEneKLM})] 
and the noninteracting PAM at $\Delta=0$ [Eq. (\ref{eq:U0disp})], regardless of the dimensionality. 
\par
If the effective mass ratio $m^*/m$ is defined by the inverse ratio of the Fermi velocity 
(slope of the dispersion relation at the Fermi level) 
to the noninteracting Fermi velocity of the hybridized band, 
the effective mass ratio does not change so much with interaction $U$ in the PAM 
as shown in Figs. \ref{fig:Akw}(f) and \ref{fig:Udep}(a). 
If the effective mass ratio is defined by the inverse ratio of the Fermi velocity 
to the noninteracting Fermi velocity of the unhybridized conduction-orbital band, 
the effective mass diverges in the small-$\delta$ limit; 
however, this is primarily due to hybridization $t_{\rm K}$ rather than interaction $U$ in the PAM. 
Without hybridization $t_{\rm K}$, the conduction-orbital band is not affected by interaction $U$ [Eq. (\ref{eq:PAM})]. 
Infinitesimal hybridization $t_{\rm K}$ can cause the divergence of the effective mass 
even without interaction $U$ [Eq. (\ref{eq:U0disp})]. 
\subsection{Doping-induced modes in the high-$|\omega|$ regime} 
The emergence of electronic states by doping is not a characteristic specific to the low-$|\omega|$ excitation. 
Electronic states can emerge even in the high-$|\omega|$ regime of $\mathcal{O}(U)$ or $\mathcal{O}(\Delta)$ in the PAM 
although doping is usually considered to only affect the properties in the vicinity of the Fermi level 
without changing those far above or below the Fermi level. 
The doping-induced electronic states far from the Fermi level can be understood 
in terms of the existence of high-energy electron-addition or electron-removal excited states 
on a doped site in a strongly correlated system (Table \ref{tbl:1site}). 
\subsection{Temperature and external-field effects} 
At nonzero temperature, 
both the hole-doped and electron-doped ground states can contribute to the spectral function of the Kondo insulator, 
which causes doping-induced modes from both the $\omega>0$ and $\omega<0$ sides in the Kondo insulating gap 
[superposition of Fig. \ref{fig:Akw}(b) and that with $\omega\rightarrow-\omega$ and $k\rightarrow\pi-k$ 
($\omega=0$ at the center of the Kondo insulating gap)], 
although the contribution is exponentially small if the temperature is much lower than the Kondo insulating gap. 
In general, the doping-induced states can appear in a spin-charge-separated insulator 
such as a Mott insulator or Kondo insulator under perturbations capable of mixing doped and undoped ground states. 
In addition, the doping-induced mode in the Kondo insulating gap would be susceptible to a magnetic field 
because it reflects magnetically excited states. 
These temperature or external-field effects can affect low-energy electronic excitation and 
might be related to the unconventional features observed in Kondo insulator materials 
\cite{SmB6dHvAScience,SmB6dHvANatPhys,SmB6dHvAiScience,YbB12dHvAJPhysCM,YbB12SdHScience,YbB12heatTransNatPhys,YbB12SdHNatPhys,YbB12HallAnomaly,YbB12SdHQM}. 
\subsection{Summary} 
The spectral features of the Kondo lattice and periodic Anderson models around Kondo insulators 
were clarified using the non-Abelian DDMRG method in one dimension, 
and the origins and properties of the characteristic modes were explained in $d$ dimensions ($d=1$, 2, and 3) 
using perturbation theory in the strong-coupling (small-$t$) regime. 
The parameter condition can be relaxed 
as long as phase transition or mode crossing does not occur. 
\par
Upon doping a Kondo insulator, an electronic mode emerges in the Kondo insulating gap. 
This mode originates from the spin excitation of the Kondo insulator, 
reflecting the spin-charge separation of the Kondo insulator 
(existence of spin excitation whose energy is lower than the charge gap). 
This characteristic does not appear in a conventional band-insulator--metal transition 
but is the same as that of the Mott transition. 
Hence, the Kondo-insulator--metal transition can be regarded as a type of Mott transition. 
\par
In the PAM, reflecting double occupancy and vacancy in localized orbitals, 
electronic modes appear in the high-$|\omega|$ regime. 
In addition to the high-$|\omega|$ modes corresponding to the upper Hubbard band 
which exist in both the presence and absence of doping, 
other high-$|\omega|$ modes emerge upon doping the Kondo insulator. 
The emergence of the high-$|\omega|$ modes as a result of doping implies that 
doping can affect not only the properties in the vicinity of the Fermi level but also those far from the Fermi level 
if hybridization with an orbital away from the Fermi level exists in strongly correlated systems. 
\par
In a hole-doped (electron-doped) Kondo insulator, 
the dominant mode for $\omega>0$ ($\omega<0$) is split into two modes in the low-energy regime by the Coulomb interaction. 
This feature is a remarkable characteristic of strongly correlated systems but has not been expected 
in conventional mean-field approximations. 
In the conventional picture, the increase of the effective mass (flattening of the dispersion relation) and 
the narrowing of the electronic-excitation gap are considered to be typical interaction effects 
on the electronic properties of a (doped) Kondo insulator. 
However, the effective mass is more sensitive to the hybridization with a localized orbital 
than to the Coulomb interaction in the PAM. 
Furthermore, the narrowing of the electronic-excitation gap in a doped Kondo insulator is not as simple as expected 
from conventional mean-field approximations. 
Instead, an electronic mode that exhibits momentum-shifted magnetic dispersion relation emerges 
in the Kondo insulating gap by doping, reflecting the spin-charge separation of the Kondo insulator. 
These innovative views on electronic excitation around a Kondo insulator would improve our understanding 
of the fundamental properties of Kondo lattice systems and heavy-fermion materials. 
\begin{acknowledgments} 
The author would like to thank S. Uji and T. Terashima for helpful discussions. 
This work was supported by JSPS KAKENHI Grant Number JP22K03477. 
Numerical calculations were partly performed on the supercomputer at the National Institute for Materials Science. 
\end{acknowledgments}

\end{document}